\documentclass[12pt]{article} 
\usepackage[latin1]{inputenc} 
\usepackage{amsmath,amssymb,amscd,epsfig}
\numberwithin{equation}{section}
\newtheorem{theorem}{Theorem} 
\numberwithin{theorem}{section}

\newtheorem{notation}[theorem]{Notation}
\newtheorem{lemma}[theorem]{Lemma}
\newtheorem{definition}[theorem]{Definition}
\newtheorem{assumption}[theorem]{Assumption}
\newtheorem{remark}[theorem]{Remark}
\newcommand{\N}{\mathbb{N}}
\newcommand{\Z}{\mathbb{Z}}
\newcommand{\R}{\mathbb{R}}
\newcommand{\C}{\mathbb{C}}
\newcommand{\calA}{\mathcal A}
\newcommand{\calI}{\mathcal I}
\newcommand{\calH}{\mathcal H}
\newcommand{\calL}{\mathcal L}
\newcommand{\calN}{\mathcal N}
\newcommand{\calM}{\mathcal M}
\newcommand{\calS}{\mathcal S}
\newcommand{\calC}{\mathcal C}
\newcommand{\OO}{\mathcal{O}}
\newcommand{\ds}{\displaystyle}
\newcommand{\wt}{\widetilde}
\newcommand{\ol}{\overline}

\newcommand{\pl}{\partial}
\newcommand{\e}{{\rm e}}
\renewcommand{\i}{{\rm i}}
\newcommand{\Le}{{\rm L}}
\newcommand{\Ha}{{\rm H}}
\newcommand{\Ce}{{\rm C}}
\newcommand{\de}{{\rm d}\,}
\newcommand{\res}{{\rm res}}
\newcommand{\sign}{{\rm sgn}}
\newcommand{\ve}{\varepsilon}
\newcommand{\vth}{\vartheta}
\newcommand{\om}{\omega}
\newcommand{\Om}{\Omega}
\newcommand{\cc}{{\rm c.c.}}
\newcommand{\boe}{{\mathbf E}}
\newcommand{\g}{\gamma}
\newcommand{\tor}{{\mathbb T}_\Gamma}

\newcommand{\whj}{{\mathcal J}}
\newcommand{\at}{\mathcal T}

\begin{document}
\title{Interaction of modulated pulses in\\
scalar multidimensional nonlinear lattices}
\author{Johannes Giannoulis\\
\small Zentrum Mathematik, TU M\"unchen\\ 
\small Boltzmannstr.~3, 
D--85748 Garching b.~M\"unchen\\ 
\small {\tt giannoulis@ma.tum.de}} 
\date{February 11, 2009}

\maketitle

\begin{abstract}
We investigate the macroscopic dynamics of sets of an 
arbitrary finite number of 
weakly 
am\-pli\-tu\-de-mo\-du\-la\-ted pulses 
in a multidimensional lattice of particles.
The latter are assumed to exhibit scalar displacement 
under pairwise, arbitrary-range, nonlinear interaction potentials 
and are embedded in a nonlinear background field.
By an 
appropriate 
multiscale ansatz, 
we derive formally 
the explicit
evolution equations for the macroscopic 
amplitudes 
up to an arbitrarily high order of the scaling parameter,
thereby deducing the resonance and non-resonance conditions  
on the fixed wave vectors and frequencies of the pulses, 
which are required for that.
%
%
%
%
The derived equations are justified rigorously 
in time intervals of macroscopic length.
%
Finally, for sets of up to three pulses we present a
complete 
list
of all 
possible 
interactions
and discuss their ramifications for the 
corresponding, explicitly given macroscopic systems.\\

\noindent{\it Key words and phrases:}  
nonlinear discrete 
lattices; 
interaction of modulated pulses; multiscale ansatz; 
derivation and 
justification of macroscopic dynamics.\\

\noindent{\it MSC 2000:} 
37K60; 
34E13, 
34C20, 
70F45, 
70K70, 
35L45, 
35L60. 
%
\end{abstract}
%
\section{Introduction}\label{s:1}

In recent years 
a big part of the research activity within 
applied mathematics has been 
focused at the investigation of so called 
\emph{multiscale problems}, cf.\ for a survey \cite{Miel06AMSM}.
%
%
One aspect of this area is to investigate for a given, mostly physical, 
phenomenon the relation between its descriptions 
at different time and/or space scales. 
A prominent paradigm 
concerns the {\it dynamics in lattices}.
Here, the underlying {\it microscopic model}
is a lattice of atoms interacting with each other, subject to some given
potential, and 
possibly embedded in a background field. 
Then, one has a complete description of the dynamics within the lattice 
by the {\it discrete system} of Newton's equations of motion for each atom
(see \eqref{M}).
However, thinking of the distance between the atoms as very small,  
one is interested in the dynamics of some  {\it macroscopic object} 
within the lattice.
For instance one could ask how initial data varying on a large space scale 
as compared to the distance of the atoms evolve in the lattice.
The corresponding dynamics would be described by a {\it macroscopic model},
which in the limit of the distance between the atoms going to zero 
is a {\it continuum system}.


Of course, this 
idea is known within the physics community
since long,
and macroscopic, or {\it effective}, continuum dynamics have been 
{\it formally derived} from
microscopic, discrete lattice systems for 
a plethora of 
phenomena,
usually by some sort of {\it  perturbation}, or {\it asymptotic}, 
{\it expansion} with respect to the 
{\it scaling parameter} $0<\ve\ll 1$ of the distance between the atoms. 
%
%

However, the derived models mostly lacked a {\it justification} in a
mathematical-analytical rigorous sense.
Moreover, at latest since the 
unexpected numerical discoveries
of Fermi, Pasta and Ulam in 1955 (see \cite{FPU55})
on the behaviour of waves 
in nonlinear 
{\it oscillator} or {\it atomic chains}, i.e., one-dimensional lattices, 
it was clear that the 
understanding of the dynamical behaviour 
of 
even one-dimensional lattices is far from complete.
This surely played a role in arising the interest also of the mathematical
community and leaded to an increased research activity regarding 
the dynamics in discrete lattices.  
We refrain here from giving an overview over the huge amount of work done
since then, and in particular the various lattices and potentials considered, 
the (macroscopic) dynamics derived, and the mathematical techniques used,
for which we 
refer the reader to the surveys \cite{CRZ05,DHR06,GHM06} 
and the references given therein.

Instead, we present in the following precisely the lattice model considered, 
the question discussed, the method used, 
and the result obtained in the present paper, 
and refer later on to literature related to them.
\\

\noindent{\bf The nonlinear lattice.} 
%
We consider the $d$-dimensional 
(Bravais) lattice $\Gamma\subset\R^d$, $d\in\N$,
\begin{equation}\label{Gamma}
\Gamma=\{\g=k_1g_1+\cdots+k_dg_d:\ k=(k_1,\ldots,k_d)\in\Z^d\},
\end{equation}
where $\{g_1,\ldots,g_d\}$ is a set of linearly independent vectors of $\R^d$
(yielding $\Gamma=\Z^d$ 
when 
$(g_i)_j=\delta_{ij}$),
%
and the \emph{microscopic model}
\begin{equation}\label{M}
\ddot x_\g(t)=\sum_{\alpha\in\Gamma}
V_\alpha^\prime\big(x_{\g+\alpha}(t){-}x_\g(t)\big)-W'\big(x_\g(t)\big),
\qquad\g\in\Gamma,\ t\in\R.
\end{equation}
This is a system of infinitely (countably) 
many coupled second-order ordinary differential equations, 
describing according to Newton's law the \emph{scalar} displacement
$x_\g(t)\in\R$ at time $t\in\R$
of a particle (e.g.\ an atom) of unit mass $1$ from its rest position 
$\g\in\Gamma$ due to the (pair-)interaction 
%
%
%
%
and on-site potentials 
$V_\alpha:\R\to\R$, $\alpha\in\Gamma$, and $W:\R\to\R$, respectively.
One can think of each mass particle of the lattice of rest position $\g$ 
as being connected to every other particle 
$\gamma{+}\alpha$, $\alpha\in\Gamma\setminus\{0\}$,  
via a spring of force $V_\alpha^\prime$, 
and of the whole lattice as being embedded in a background (external) 
force field $W^\prime$.
Since the interaction and on-site potentials $V_\alpha$ and $W$ 
are independent of the position of the particles, and 
since all particles have equal mass, we speak of a 
\emph{monoatomic, homogeneous lattice}.
%
Moreover, 
since naturally 
$V_\alpha^\prime(x_{\g+\alpha}{-}x_\g)
=-V_{-\alpha}^\prime(x_\g{-}x_{\g+\alpha})$,
we assume
$V_\alpha(x)=V_{-\alpha}(-x)$ for all $\alpha\in\Gamma$, $x\in\R$.
%
%
Note that we allow for interactions among \emph{pairs} of particles, 
which are arbitrarily far apart from each other. 
%
In order for our results to hold true, 
we will need to impose some rather strong decay conditions 
(cf.\ \eqref{condalpha})
on the interaction potentials $V_\alpha(x)\to 0$ as $|\alpha|\to\infty$, 
which however seem physically plausible.    
%
Of course, these conditions 
are satisfied
if we assume that pairs of particles interact only for distances up to some 
finite range $R>0$, i.e., $V_\alpha=0$ for $|\alpha|>R$. 
For instance, for $\Gamma=\Z$ 
and $V_\alpha=0$ if $|\alpha|>1$ we obtain the
classical 
oscillator chain with only nearest-neighbour interactions 
\begin{equation}\label{chain}
 \ddot x_\g(t)=V_1^\prime\big(x_{\g+1}(t){-}x_\g(t)\big)
 -V_{1}^\prime\big(x_\g(t){-}x_{\g-\alpha}(t)\big)
 -W^\prime\big(x_\g(t)\big),
 \qquad\g\in\Z,\ t\in\R,
\end{equation}
which for $W=0$ and anharmonic $V_1$, $ V_1(x)\neq a_{1,1}\frac{x^2}2$, 
is the 
Fermi-Pasta-Ulam (FPU) chain, 
and for harmonic $V_1$ and anharmonic $W$ 
the so called Klein-Gordon (KG) 
chain.
%
%

Throughout this paper we assume $V_\alpha,\,W\in\Ce^{N+2}(\R)$ for some 
$N\in\N$ to be specified, 
and $V_\alpha(0)=V^\prime_\alpha(0)=W(0)=W^\prime(0)=0$,
%
which allows for the Taylor expansions
\begin{equation}\label{VW}
\begin{cases}
\ds V^\prime_\alpha(x)
=\sum_{n=1}^{N} a_{n,\alpha}x^n+V^\prime_{N+1,\alpha}(x),
\qquad V^\prime_{N+1,\alpha}(x):=
\frac{V^{(N+2)}_\alpha(\xi x)}{(N{+}1)!}x^{N+1},
\\[5mm]
\ds W^\prime(x)
=\sum_{n=1}^{N} b_n x^n+W^\prime_{N+1}(x),
\qquad W^\prime_{N+1}(x):=
\frac{W^{(N+2)}(\kappa x)}{(N{+}1)!}x^{N+1}
\end{cases}
\end{equation}
with $a_{n,\alpha}:=V^{(n+1)}_\alpha(0)/(n!)$, $b_n:=W^{(n+1)}(0)/(n!)$
and $\xi=\xi_{N,\alpha}(x),\ \kappa=\kappa_N(x)\in(0,1)$.
Note, that $V_\alpha(x)=V_{-\alpha}(-x)$ implies 
$a_{n,\alpha}=({-}1)^{n+1}a_{n,-\alpha}$. 
%
By \eqref{VW}, 
the microscopic model \eqref{M} reads equivalently
\begin{align}\label{Mexpl}
\ddot x_\g
&=
\sum_{n=1}^{N}
\left(\sum_{\alpha\in\Gamma} a_{n,\alpha}(x_{\g+\alpha}{-}x_\g)^n
-b_nx_\g^n\right)
+\sum_{\alpha\in\Gamma}V^\prime_{N+1,\alpha}(x_{\g+\alpha}{-}x_\g)
-W^\prime_{N+1}(x_\g).
\end{align}
%
%
%

Finally, note that \eqref{M} is an infinite-dimensional Hamiltonian system 
  with conserved total energy (Hamiltonian) 
  ${\cal H}(x,\dot x)={\cal K}(\dot x)+{\cal U}(x)$, 
and kinetic and potential energies
  \begin{equation}\label{KU}
  {\cal K}(\dot x)=\frac12\sum_{\g\in\Gamma} \dot x_\g^2,
  \qquad
  {\cal U}(x)
  =\sum_{\g\in\Gamma}
  \left(\frac12\sum_{\alpha\in\Gamma}V_\alpha(x_{\g+\alpha}(t){-}x_\g(t))
  +W(x_\g(t))\right).
  \end{equation}\\



\noindent{\bf Pulses and their interactions.} 
The linearized microscopic model \eqref{M} 
\begin{equation*}
\ddot x_\g
=\sum_{\alpha\in\Gamma}a_{1,\alpha}(x_{\g+\alpha}{-}x_\g)-b_1 x_\g
\end{equation*}
possesses the plane-wave solutions or \emph{pulses}
$x_\g(t)=\boe(t,\g)+\cc$, where $\cc$ denotes the complex conjugate of 
$\boe(t,\g):=\e^{\i(\om t+\vth{\cdot}\g)}$,
provided the \emph{frequency} $\om\in\R$ 
and the \emph{wave vector} $\vth\in\tor$
satisfy the \emph{dispersion relation}
\begin{equation}
\label{DR}
\om^2=\Om^2(\vth)
:=\sum_{\alpha\in\Gamma}a_{1,\alpha}\left(1{-}\e^{\i\vth{\cdot}\alpha}\right)
+b_1
=\sum_{\alpha\in\Gamma}a_{1,\alpha}\left(1{-}\cos(\vth{\cdot}\alpha)\right)
+b_1
\end{equation}
(the latter equality follows from $a_{1,\alpha}=a_{1,-\alpha}$, see above).
%
Here, $\tor:=\R^d_\ast/_{\Gamma_\ast}$ is the $d$-dimensional (dual) torus
associated to the lattice $\Gamma$, where $\R^d_\ast=\mathrm{Lin}(\R^d)$ is
the dual of $\R^d$ and 
$\Gamma_\ast:=\{\vth\in\R^d_\ast:\ 
\vth:\alpha\mapsto\vth\cdot\alpha,\  
\vth\cdot\alpha\in 2\pi\Z\ \forall\ \alpha\in\Gamma\}$
is the dual lattice to $\Gamma$. 
%
In the following 
we assume that the \emph{stability condition} 
\begin{equation}
\label{SC}
\Om^2(\vth)>0\quad\text{for all  $\vth\in\tor$}
\end{equation}
is satisfied, 
and set the \emph{dispersion function} $\Om(\vth)>0$ 
for all  $\vth\in\tor$.
Note, that 
\eqref{SC} implies that there are no pulses with frequency $\om=0$,
and in particular that 
$\boe(t,\g)=1$ is not a pulse.
%
The stability condition can be guaranteed by
appropriate choice of the coefficients $a_{1,\alpha}$ and $b_1$ 
of the harmonic parts of the potentials $V_\alpha$ and $W$.
For instance, it is surely satisfied if we assume 
$\ds\min\{b_1, 
2\sum_{\alpha\in\Gamma}\min\{a_{1,\alpha},0\}
+b_1\}>0$.

The dispersion relation enables the following characterization of pulses:
\emph{A pair 
$(\vth,\om)\in\tor\times\R$
represents 
a pulse 
if and only if $\delta:=\Om^2(\vth)-\om^2=0$.}
The pair $-(\vth,\om)$ 
represents the same pulse,
while $\pm(\vth,-\om)$ represent the associated pulse traveling 
in opposite direction. 
In the following we will denote pulses mostly by the 
pairs 
representing them. 
In fact, generalizing the use of this notation, \emph{we will denote arbitrary
functions $\boe(t,\g)=\e^{\i(\om t+\vth{\cdot}\g)}$ by the pair 
$(\vth,\om)\in\tor\times\R$}.  
Thus, be aware that this notation does not imply that $\boe$ is a pulse,
unless the corresponding pair satisfies the dispersion relation \eqref{DR}.\\

In this paper
we will deal with sets of 
\emph{$\nu\in\N$ different, fixed pulses} 
$\{(\vth_j,\om_j): j=1,\ldots,\nu\}$, 
such that $\delta_j:=\Om^2(\vth_j)-\om_j^2=0$ 
and $\boe_j:=\e^{\i(\om_j t+\vth_j{\cdot}\g)}\neq \boe_i$ for $j\neq i$.
With the notation $(\vth_{-j},\om_{-j}):=-(\vth_j,\om_j)$ this set of pulses
can be written equivalently as $\{(\vth_j,\om_j): j\in\calN\}$,
where $\calN:=\{-\nu,\ldots,-1,1,\ldots,\nu\}$, and it holds 
$\boe_{-j}=\ol{\boe_{j}}$, $\ol{z}$ denoting the complex conjugate ($\cc$)
of $z\in\C$.

Since the microscopic 
model \eqref{M} is nonlinear, 
\emph{products of pulses} 
$\boe_{(j_1,\ldots,j_k)}:=\boe_{j_1}{\cdots}\boe_{j_k}$
with 
$(j_1,\ldots,j_k)\in{\calN}^k$ 
will play an essential role throughout 
this paper.
%
%
However, 
\emph{different index-vectors $(j_1,\ldots,j_k)$ can yield the same 
product.} 
(Confer for instance the products corresponding to the indices 
$(p,q)$, $(q,p)$, or $(p,-p,q)$, $(-p,p,q)$, $(p,q,-p)$, $(-p,q,p)$, 
$(q,p,-p)$, $(q,-p,p)$ for $p,q\in\calN$.) 
Moreover, \emph{the product for 
$(j_1,\ldots,j_k)$ can
equal a product for 
$(i_1,\ldots,i_\kappa)$ 
with $\kappa\neq k$}.
(Confer for instance the products corresponding to the indices 
$(p,-p,q)$ and $q$.) 
%
%
Hence, in order 
to identify each appearing product of pulses by a unique index, 
we introduce the following
\begin{notation}\label{pulseproduct}\ \\
{\bf (a)} 
We denote by $\whj_m$ a representant of all those 
$(j_1,\ldots,j_\kappa)\in{\calN}^\kappa$, $\kappa\le k$, 
which lead \linebreak\phantom{{\bf (a)}}  to the same product 
$\boe_{\whj_m}(t,\g)
:=\e^{\i(\om_{\whj_m}t+\vth_{\whj_m}\cdot\g)}
:=\boe_{(j_1,\ldots,j_\kappa)}(t,\g)$,
where $\vth_{\whj_m}:=$ \linebreak\phantom{{\bf (a)}}
$\vth_{j_1}{+}\ldots{+}\vth_{j_\kappa}$ in $\tor$
and
$\om_{\whj_m}:=\om_{j_1}{+}\ldots{+}\om_{j_\kappa}$.
The set of all (different) re\-pre\-sent- \linebreak\phantom{{\bf (a)}}  
ants $\whj_m$  
within 
$\bigcup_{\kappa=1}^k{\calN}^\kappa$ is
denoted by $\at_k$.
The index $m$ denotes $\whj_m\in{\calN}^m\setminus\at_{m-1}$ 
with \linebreak\phantom{{\bf (a)}}
$\whj_1:=j\in\calN=\at_1$.\\
%
%
{\bf (b)} The representation $\boe_{\whj_m}=\boe_{(j_1,\ldots,j_\kappa)}$
is denoted by $\whj_m=(j_1,\ldots,j_\kappa)$.
More generally, \linebreak\phantom{{\bf (b)}} 
$\boe_{(j_1,\ldots,j_k)}=\boe_{
(i_1,\ldots,i_\kappa)
}$ 
is denoted by $(j_1,\ldots,j_k)=
(i_1,\ldots,i_\kappa)
$.
%
%
%
\end{notation}
%
(
According to this notation,
the products $\boe_{(p,q)}=\boe_{(q,p)}$ from above have the same 
representant $\whj_m=(p,q)=(q,p)$ with $m=1$ iff $(p,q)=j\in\calN$ and $m=2$
iff $(p,q)\neq j$ for all $j\in\calN$. Moreover, 
$\boe_{(p,-p,q)}=\boe_q$ has the representant $\whj_1=q\in\calN$, as do all
permutations of the indices within $(p,-p,q)$.)
%
%

By use of Notation \ref{pulseproduct}, we can state concisely 
some observations on 
sets of pulses.
%
\begin{remark}\label{rempuls}
Let $\{(\vth_j,\om_j): j\in\calN\}$, 
$\calN=\{\pm 1,\ldots,\pm\nu\}$,
be a set of 
$\nu$
pulses. Then
\begin{enumerate}
\item\label{obs1}
It always holds $\calN^{k-2}\subset\calN^{k}$, since 
$(j_1,\ldots,j_{k-2})=(j_1,\ldots,j_k,j,-j)$ for any $j\in\calN$.
In particular, $(j_1,\ldots,j_k)\in{\calN}^k$ can have a representant 
$\whj_m\in\at_{k-2}$. (Cf., e.g., for $k=3$ the example $(p,-p,q)$ above.) 
\item\label{obs2}
In general $\calN^{k-1}\not\subset\calN^{k}$.  
E.g., for $\calN=\{\pm1\}$, we get $\calN^2=\{\pm(1,1),\pm(1,-1)\}$ 
and $\at_3\cap\calN^3=\{\pm1,\pm(1,1,1)\}$.
Since $\boe_{\pm 1}\neq\boe_{1}^2\neq\boe_{\pm1}^3$ 
and $\boe_{\pm 1}\neq\boe_1\boe_{-1}=1\neq\boe_{\pm1}^3$,
we obtain $\calN\cap\calN^2=\emptyset$ and $\calN^2\cap\calN^3=\emptyset$,
while $\calN\subset\calN^3$, in line with Obs.\ \ref{obs1}.
\item\label{obs3}
If for each $j\in\calN$ there exist $p,q\in\calN$ such that $(p,q)=j$,
then $\calN^{k-1}\subset\calN^{k}$, 
since $(j_1,\ldots,j_{k-2},j)=(j_1,\ldots,j_{k-2},p,q)$.
In this case $\at_k\subset\calN^k$.
\end{enumerate}
\end{remark}
%

In general, a product of $k$ pulses 
$
\boe_{j_1}\ldots\boe_{j_k}$ 
need not be a pulse.
However, if it is a pulse, 
we make the following definition.
\begin{definition}\label{resonance}
$k\in\N$ pulses $(\vth_{j_i},\om_{j_i})$, $i=1,\ldots,k$, 
are \emph{in resonance (of order $k$)} or 
\emph{interacting with each other (in order $k$)} if 
\begin{equation*}
(\om_{j_1}{+}\ldots{+}\om_{j_k})^2=\Om^2(\vth_{j_1}{+}\ldots{+}\vth_{j_k})
\end{equation*}
or, equivalently, by use of Notation \ref{pulseproduct}, if 
\begin{equation}\label{rescond}
\delta_{\whj_m}:=\Om^2(\vth_{\whj_m})-\om_{\whj_m}^2=0,
\end{equation}
where $\whj_m=(j_1,\ldots,j_k)$.
\end{definition}
%
Finally, we conclude our discussion of pulses, their products and their 
interactions, 
by providing a definition which will be essential for our results.
\begin{definition}\label{defclosure}
The set of pulses $\{(\vth_j,\om_j):\ j\in\calN\}$
is \emph{closed under interactions up to order $k$} 
(or for short:  \emph{$k$-closed}) 
if 
\begin{equation}\label{closedness}
\delta_{\whj_m}:=\Om^2(\vth_{\whj_m})-\om_{\whj_m}^2\neq 0
\quad\text{for $\whj_m\in\at_k\setminus\calN$.} 
\end{equation}
\end{definition}
In other words, a set of pulses 
satisfies the \emph{closedness condition} \eqref{closedness}
if 
\emph{none of the products of up to $k$ (possibly identical) pulses of the set
are \emph{new} pulses}, i.e., pulses which are not included in the set.
As 
the negation of
\eqref{rescond}, the conditions  \eqref{closedness} are
refered to also as \emph{nonresonance conditions}.\\
%
%

%
%

\noindent{\bf Macroscopic dynamics of modulated pulses - Formal derivation.}
For a given pulse $\boe(t,\g)=\e^{\i(\om t+\vth\cdot\g)}$  
an \emph{(amplitude-)modulated pulse} 
is a function of the form 
\begin{equation*}
x_\g(t)=\ve^a A(\ve^b t,\ve(\g{-}ct))\boe(t,\g)+\cc
\end{equation*}
with the \emph{scaling parameter} $0<\ve\ll1$
and the \emph{amplitude} or \emph{envelope} $(\ve^a)A:\R\times\R^d\to\C$.
The latter depends on the {\it macroscopic space variable}
$y=\ve(\g-ct)$, where we allow for possibly moving-frame space coordinates of 
velocity $c\in\R^d$.
The dependance of $A$ on such a $y$ means that the amplitude varies at a much
larger space scale of order $\OO(\frac1\ve)$ 
--- or, for space-periodic $A$, that it has a much greater wavelength --- 
in comparison to the space (wavelength) scale of order $\OO(1)$ 
of $\boe(0,\g)=\e^{\i\vth\cdot\g}$, since  $0<\ve\ll1$.
The choice of the powers $a$ and $b$ determines the 
size of the amplitude and the length of its 
time scale $\tau=\ve^b t$. 
%
This choice is motivated by the physical phenomena one wants to study, and
shapes the macroscopic equations obtained, 
cf.\ the discussion of related literature below.
%
%

Here, for a set of $\nu\in\N$ different, fixed pulses 
$\{(\vth_j,\om_j)\,:\, j\in\calN\}$
we consider their small-amplitude macroscopic modulations
\begin{equation}\label{X1A}
(X^{A,\ve}_1)_\g(t):=\ve \sum_{j=1}^\nu A_j(\ve t,\ve\g)\boe_j(t,\g)+\cc
\end{equation} 
with hyperbolically scaled amplitudes 
$A_j:\R{\times}\R^d\to\C$:
$\tau=\ve t$, $y=\ve j$.
We are interested in the {\it macroscopic dynamics} of such modulated pulses
in \eqref{M}.
This means that, 
making 
for solutions to 
the \emph{microscopic model} 
\eqref{M} 
the 
\emph{multiscale ansatz}
\begin{equation}\label{multiscale} 
x_\g(t)=(X^{A,\ve}_1)_\g(t)+\OO(\ve^2),
\end{equation} 
we are interested in the dynamics of the amplitudes $A_j$,
%
%
or more generally, making the 
ansatz
\begin{align}
x_\g(t)=(X^{A,\ve}_N)_\g(t)+\OO(\ve^{N+1}),
\quad
\label{XNA}
X_N^{A,\ve}:=\sum_{k=1}^N \ve^k
\sum_{\whj_m\in\at_k}
A_{k,\whj_m}\boe_{\whj_m}
\end{align} 
for $N\in\N$, 
in the macroscopic dynamics of the 
functions
$A_{k,\whj_m}:\R{\times}\R^d\to\C$, $(\tau,y)\mapsto A_{k,\whj_m}(\tau,y)$,
(with $\overline{A_{k,\whj_m}}=A_{k,-\whj_m}$ 
in order to obtain $X_N^{A,\ve}\in\R$),
and $A_{1,j}=A_j$ for $j\in\calN$  
(recall the definition
of 
$\whj_m$ 
in Notation \ref{pulseproduct}).
%
%
Note here, that in view of Remark \ref{rempuls}, Obs.\ \ref{obs2},
for each $k$ we take into account $\whj_m\in\at_k$ instead 
of $\whj_m\in\calN^k$ 
(cf.\ also Sec.\ \ref{singleclosed3}).


%

Inserting  \eqref{XNA} into the microscopic model  \eqref{M}
and carrying out the operations,
we obtain on the left- and right-hand sides
expansions of the form \eqref{XNA},
where,
in particular due to the nonlinearity of \eqref{M},
the macroscopic coefficients are products of functions $A_{\kappa,\whj_m}$,
$\kappa\le k$ 
their space and time derivatives.
%
%
%
Thus, 
in order for the microscopic system \eqref{M} to be satisfied 
up to order $\ve^N$
by the approximation \eqref{XNA},
and 
since at each $k$ the microscopic patterns $\boe_{\whj_m}$ are different from
each other for $\whj_m\in\at_k$ and are nonvanishing (of modulus $1$),
the macroscopic coefficients on the left- and right-hand sides of the
corresponding microscopic patterns have 
necessarily 
to be equal.
%
%
This yiels an hierarchy of equations for the functions 
$A_{k,\whj_m}$ in \eqref{XNA}, 
which guarantee that $X_N^{A,\ve}$ satisfies \eqref{M} up to order $\ve^N$,
and constitutes the {\it formal derivation} of their macroscopic dynamics,
carried out in Section \ref{formalderiv}.
%
%
In particular, due to the hyperbolic scaling $\tau=\ve t$, $y=\ve\g$ and the
scaling by $\ve$ of the first-order amplitudes $A_j$, 
it turns out that the latter are determined by the equations for
$\ve^2\boe_j$, which contain the macroscopic first-order time- and 
space-derivatives $\partial_\tau A_j$ and $\nabla_y A_j$,
and, due to the cubic terms in the potentials $V_\alpha,W$, 
the products $A_pA_q$ for which $\boe_p\boe_q=\boe_j$,
whenever the pulses $(\vth_p,\om_p)$ and  $(\vth_q,\om_q)$  
interact to generate the pulse $(\vth_p,\om_p)+(\vth_q,\om_q)=(\vth_j,\om_j)$,
cf. \eqref{macro1}. 
Note here, that due to the underlying hyperbolic scaling and the scaling of
the amplitudes by $\ve$, coupling of more 
than two pulses is not shown in the leading order effective dynamics.
Moreover, as we will see in more detail in Section \ref{formalderiv}, 
in order to calculate the functions 
$A_{k,\whj_m}$, $k\le N$, $\whj_m\in\at_N\setminus\calN$, 
we have to require 
that the system is closed under interactions up to order $N$.\\
%
%


\noindent{\bf Justification.}
The macroscopic evolution equations we obtained by the above procedure of the
formal derivation, however, establish only the {\it necessary} conditions on 
$A_{k,\whj_m}$, which solutions to the microscopic model \eqref{M} have to
satisfy {\it if} they are of the form \eqref{XNA}.
 The existence of such solutions is not at all guaranteed. 
 Indeed, there exist counterexamples, cf. \cite{Sch95,Sch05}.

 Since for $t=0$ we can prescribe the initial data in such a way 
 that they have the form \eqref{XNA},
 the question of the {\it justification} (or the {\it validity}) 
 of the formally derived macroscopic equations is answered in the affirmative,
 if we can show that solutions to the microscopic model 
 \eqref{M} with such initial data maintain this form also on time intervals 
 $[0,\tau_0]$ of positive {\it macroscopic} length $\tau_0>0$.
 The reason why such a condition should be satisfied on the 
 macroscopic time scale $\tau=\ve t$ is obvious, 
 considering that the scaling parameter is very small, $0<\ve\ll1$.  
%

 For the systems considered in the present paper, 
 the justification of the macroscopic equations 
 derived formally in Section \ref{formalderiv}, 
 is carried out in Section \ref{jf}.
%
%
 In particular, the justification result, Theorem \ref{justiftheorem},
 states that if the error between the approximation 
 $X_{N-1}^{A,\ve}$ in \eqref{XNA} 
 {\it obtained by the solutions} $A_{k,\whj_m}$ 
 {\it to the macroscopic equations} 
 (viz.\ \eqref{macro1}, \eqref{macro2} for $N=2$ 
 and \eqref{macro3}, \eqref{macro4} for $N\ge 3$) 
 and a solution of the microscopic model \eqref{M}
 is initially of order $\ve^\beta$ with $\beta\in(1,N-d/2]$, 
 then it will remain in the same order for times $t\le \tau_0/\ve$,
 $\tau_0>0$ restricted only by the existence of solutions of the macroscopic
 equations.
 In particular this is the order of the error when the initial data for the
 microscopic system \eqref{M} 
 are given by 
 $X_{N-1}^{A,\ve}$.
 Thus, naturally, the higher the order of the approximation the smaller its 
 error with respect to a true solution.
 However, note that there is a lower bound for the order $N$ in the ansatz 
 \eqref{XNA} needed in order to get reasonable results, 
 depending on the space dimension $d$ of the lattice 
 and the scaling by $\ve$ of the amplitude.
 Hence, only in the case of a one-dimensional lattice, 
 i.e.\ an oscillator chain, 
 the order $N=2$
 is sufficient to obtain {\it valid} macroscopic dynamics 
 of the first-order amplitudes.
 For the other physically relevant space dimensions $d=2$ or $d=3$ we need 
 to use $N=3$ in \eqref{XNA} in order to be able to determine the full 
 second-order approximation $X_2^{A,\ve}$ 
 which gives the valid macroscopic dynamics.  
%
%



The proof of Theorem \ref{justiftheorem} consists on 
a Gronwall argument for the {\it error} $R_\ve=\ve^{-\beta}(x-X_N^{A,\ve})$
between approximation and original solution $x$ of \eqref{M} 
in the energy norm of the microscopic system. 
This needs, on the one hand, an estimate of the
residual terms $\res\big(X_N^{A,\ve}\big)=\OO(\ve^{N+1})$ 
in the in the $\ell^2(\Gamma)$-norm.
Since they depend on the solutions $A_{k,\whj_m}$ of the macroscopic equations,
we obtain due to rescaling 
$\|\res\big(X_N^{A,\ve}\big)\|_{\ell^2}=\OO(\ve^{N+1-d/2})$ 
provided the regularity of $A_{k,\whj_m}$ is sufficiently high.
On the other hand, we have to estimate the difference between 
the nonlinear terms of the original solution $x$ and of the approximation 
$X_N^{A,\ve}$
in the integral formulation of the differential equation for $R_\ve$,
see \eqref{varconst}.
Since they are quadratic, they can be matched by the hyperbolic scaling of the
system where $\tau=\ve t$, which is essential for the Gronwall argument,
the latter yielding finally the justification of the macroscopic equations for 
$\tau\le\tau_0$.

The strategy of the proof is classical within the theory of modulation
equations and not restricted to discrete systems. For a concise presentation
in the case of continuum systems, see \cite{KSM92}. 
For a further example on how essential the matching
between macroscopic time scale and nonlinearity is, 
cf.\ \cite{GM04,GM06}, where the
dispersive scaling $\tau=\ve^2 t$, $y=\ve(\g -\Om^\prime(\vth)t)$ was used 
in order to derive the nonlinear Schr\"odinger equation for an oscillator
chain. 
As long as the nonlinearities of the system are 
cubic, \cite{GM04}, the proof is straightforward.
In order to include cubic potentials, \cite{GM06}, one has first to apply a
normal-form transformation on the system. 
In the present case, due to the hyperbolic scaling, 
the latter is not needed.\\
%

\noindent{\bf Examples.}
We conclude our paper with a list of examples in Section \ref{secexamples}.
In particular, we classify all possible interactions (or order $2$) for up to $\nu=3$ pulses 
which form systems closed under interactions up to order $N=2$, and give the
corresponding macroscopic dynamics. 
The main purpose of the section is on the one hand to show how the resonances 
are mirrored in the macroscopic dynamics and on the other hand to clarify the
role of the closedness condition. The main observation is that if a system 
is not closed, i.e.\ if the considered pulses generate {\it new} pulses not 
taken into account by the multiple scale ansatz \eqref{XNA} then 
the macroscopic equations describe only the dynamics in trivial cases.
%
In other words, in order 
to be able to detect the dynamics of
interacting pulses macroscopically {\it one has to take into account all 
pulses {\em generated} by the considered system}. 
However, {\it generating} pulses can of course be ignored. 

As explained above, for one-dimensional lattices the requirement on the
systems of pulses to be closed under interactions up to order $N=2$ is
sufficient in order to obtain valid macroscopic dynamics.
For two- and three-dimensional lattices the systems need to be closed under
interactions up to order $N=3$. This increases the number of nonresonance
conditions \eqref{closedness}
which have to be satisfied 
in order to obtain the second-order approximation $X_2^{A,\ve}$ needed 
for the justification result.
However, it leaves the resonance conditions 
and thus the macroscopic equations for the first-order amplitudes unchanged. 
Exemplarily, we give the 
macroscopic equations for the
second-order amplitudes $A_{2,j}$ $j=1,\ldots,\nu$ for a single pulse
(Sec.\ \ref{singleclosed3}) 
and for the three-wave-interaction (Sec.\ \ref{433}). 
%

Finally, in Section \ref{interactionexistence}
%
we look closer at the resonance condition for a three-wave-interaction in a
one-dimensional lattice with only nearest-neighbour interaction 
and a stabilizing on-site potential.
In particular, we determine the coefficients of the harmonic part of the
interaction potential which allow for the existence of 
three-wave-interactions. 
It turns out that indeed this is possible only for a
small range of repulsive harmonic parts.

For a more precise discussion and interpretation of the phenomena observed in
the interaction of pulses, see the introduction to Sec.\ \ref{secexamples}.
\\
%

\noindent{\bf Related literature.}
The present work is closest related to work which also uses the modulational
approach described above in order to justify modulation equations derived 
formally by asymptotic expansions in the scaling parameter.
For a general description of the method in the case of lattices, 
see \cite{GHM06}.
As already mentioned, this method was used also in \cite{GM04,GM06}, 
justifying the nonlinear Schr\"odinger (nlS) equation describing 
for a single pulse in an oscillator chain the deformation of its 
(small) amplitude in the macroscopic variables 
$\tau=\ve^2 t$, $y=\ve(\g-\Om^\prime(\vth)t)$, 
$\Om^\prime(\vth)$ the microscopic group velocity of the carrier plane wave.
In this case the dispersive scaling, i.e.\ in particular the longer time scale
$\tau=\ve^2 t$,  was chosen in order to allow for the amplitude scaled be $\ve$
(and corresponding to a weak nonlinearity) to deform, 
and as a consequence the nlS equation was justified for $t\in[0,\tau_0/\ve^2]$.
Under the hyperbolic scaling used here, 
we observe only the transport of the amplitude, see \eqref{singletransport}.
%

As indicated previously, the modulational approach is not at all restricted to
lattice systems. In the contrary, 
it has been previously applied to continuous systems,
see for an overview \cite{Kal89,Sch05} 
and for a short exposition of the idea  \cite{KSM92}.
Concerning the interaction of pulses, 
in \cite{SW03} it is shown that resonating water waves which are subject to 
weak surface tension can be approximated by using the same scaling as in 
\eqref{XNA}, by a system of three-wave-interaction equations of the form 
\eqref{3wi}.
For a discussion of this 
system and its various
physical applications, see \cite{BS90,Kau80,Kea99,KRB79,SW03}
and the references given therein. 
Moreover, we would like to point out the structural similarities of the
derived three-wave interaction equation \eqref{3wi} 
and its underlying setting 
with the Boltzmann-like equation describing the collision of phonons in the
kinetic limit 
and its relations to wave turbulence,
see \cite{Spo06}. 

In \cite{SW00} a coupled system of Korteweg-de Vries
equations was justified as governing the evolution of the amplitudes $f$ and
$g$ of two counter-propagating waves in an FPU chain in the "small,
long-wavelength limit" (cf.\ also \cite{FP99}), i.e., making the ansatz
\begin{equation*}
x_\g(t)=\ve^2 f(\ve(\g+ct),\ve^3 t)+\ve^2 g(\ve(\g-ct),\ve^3 t)+\OO(\ve^4).
\end{equation*}
While the previous ansatz contains no internal microsctructure, 
in \cite{SUW09} the interaction of two (weakly) amplitude modulated pulses of 
different group velocities $c_1\neq c_2$ 
and 
{\it time-independent amplitudes} is considered in the chain \eqref{chain}
under the dispersive scalings $\tau=\ve^2 t$ $y=\ve(\g-c_{1,2}t)$,
and it is proved 
that after interaction the amplitudes retain their shape but experience a
shift of order $\OO(\ve)$ in position;
%
%
see also \cite{BF06}. 
Analogous results are obtained in \cite{CBea07,CBea08} on a continuous
one-dimensional string.

In a different physical setting but with the closely related WKB-approximation
approach, in \cite{GMS08} the amplitude equations for interacting modulated
pulses of a nonlinear Schr\"odinger equation with periodic potential 
are justified in the semiclassical scaling.
%

The results obtained in the present paper rely for the derivation 
of the macroscopic equations fully on Newton's equations of
motion \eqref{M}. However, this microscopic model possesses Lagrangian
and Hamiltonian structure (LHS), cf.\ \eqref{KU}. 
In \cite{GHM08a} a general framework is presented for the {\it direct} 
reduction from the microscopic LHS of the macroscopic LHS corresponding to
the limit equation by use of the related two-scale ansatz.
Among several examples (with or without microstructure) 
this is also performed for the three-wave interaction of 
Sec.\ \ref{threepulsesclosedN2}, Case \ref{case3wi} (see also \cite{GHM08b}).  

We conclude our (non-exhaustive) survey of related literature by 
recalling three features of the setting we discuss in the following: first, 
the assumption of scalar displacement $x_\g\in\R$, second, the existence of a
stabilizing on-site potential, and, third, the smallness of the amplitudes
(weakly-nonlinear regime).
These features imply also possible generalizations. 
In this direction, 
\cite{Mie06} contains a thorough analysis of macroscopic continuum limits 
in multidimensional {\it linear} lattices, while \cite{DHM06,DHR06} 
present first results concerning the Whitham modulation equation as the
macroscopic limit in oscillator chains in the fully nonlinear case.\\

The paper is organized as indicated by the introduction: in the following 
Section \ref{formalderiv} we derive formally the macroscopic equations for the
functions $A_{k,\whj_m}$ of a general approximation \eqref{XNA}, 
which are then justified in Section \ref{jf}. In Section \ref{secexamples}
we present all possible macroscopic systems (mainly) for the first-order
approximations for systems of up to $\nu=3$ pulses, 
discuss exemplarily the significance of the closedness
condition, and prove in a typical case the existence of interacting pulses 
in oscillator chains. The main observations are summarized in the beginning of
the section.















\section{Formal derivation}\label{formalderiv}
In this section we derive the macroscopic equations which the functions 
$A_{k,\whj_m}$ of the multiscale ansatz $X_N^{A,\ve}$, see \eqref{XNA}, 
have to satisfy necessarily in order for the microscopic
model \eqref{M} or, equivalently, \eqref{Mexpl} to be satisfied 
up to terms of order $\ve^N$.
We follow the procedure outlined in the Introduction. 
We chose to present the derivation in full detail,
in order to enable the interested reader to see clearly 
the origin and structure of the resulting macroscopic equations 
in their general form.
However, the reader interested only in the equations themselves may proceed
directly to equation \eqref{onwards}, which summarizes the obtained results.
We present the macroscopic equations explicitly in order to provide 
them for any possible further use, as for instance for the derivation 
of the macroscopic equations in the concrete examples of 
Section \ref{secexamples}.

We start by inserting the ansatz $X_N^{A,\ve}$, see \eqref{XNA}, 
into the microscopic model \eqref{Mexpl},
\begin{align}
\ddot{X}_N^{A,\ve}
=&
\sum_{n=1}^{N}\Big(\sum_{\alpha\in\Gamma}a_{n,\alpha}
\big((X_N^{A,\ve})_{\cdot+\alpha}{-}X_N^{A,\ve}\big)^n
-b_n\big(X_N^{A,\ve}\big)^n\Big)
\notag\\&
+\sum_{\alpha\in\Gamma}
V^\prime_{N+1,\alpha}\big((X_N^{A,\ve})_{\cdot+\alpha}{-}X_N^{A,\ve}\big)
-W^\prime_{N+1}\big(X_N^{A,\ve}\big),
\label{insertedansatz}
\end{align}
where $(x_{\cdot+\alpha})_\g:=x_{\g+\alpha}$ for $x\in\ell^2(\Gamma)$.
Next, we expand the left- and right-hand sides of \eqref{insertedansatz}
with respect to terms of the form  $\ve^k\boe_{\whj_m}$, $k=1,\ldots,N$, 
$\whj_m\in\at_k$
(cf.\ Notation \ref{pulseproduct}).
For the left-hand side one has
\begin{align}
\ddot{X}_N^{A,\ve}
&
=\sum_{k=1}^N\ve^k\sum_{\whj_m\in\at_k}
\left(\ve^2\partial_\tau^2 A_{k,\whj_m}
+\ve2\i\om_{\whj_m}\partial_\tau A_{k,\whj_m}
-\om_{\whj_m}^2 A_{k,\whj_m}\right)
\boe_{\whj_m}
\notag\\&
=\sum_{k=1}^{N+2}\ve^kt_{k}
,\qquad t_k
:=\sum_{\whj_m\in\at_k}
\left(\partial_\tau^2 A_{k-2,\whj_m}
+2\i\om_{\whj_m}\partial_\tau A_{k-1,\whj_m}
-\om_{\whj_m}^2 A_{k,\whj_m}\right)\boe_{\whj_m},
\label{tk}
\end{align}
where $A_{k,\whj_m}=0$ for $k\in\Z\setminus\{1,\ldots,N\}$
or $m>k$.
For the expansion of the right-hand side of \eqref{insertedansatz} 
we first expand $(X_N^{A,\ve})_{\cdot+\alpha}{-}X_N^{A,\ve}$.
Since 
$(A_{k,\whj_m}\boe_{\whj_m})_{\cdot+\alpha}{=}
A_{k,\whj_m}({\cdot},{\cdot}{+}\ve\alpha)\e^{\i\vth_{\whj_m}\cdot\alpha}
\boe_{\whj_m}$,
we use the Taylor expansion of $A_{k,\whj_m}$ with respect to $y$,
assuming 
for the moment  
that $A_{k,\whj_m}(\tau,\cdot)\in\Ce^{N-k+1}(\R^d)$ for $\tau\in[0,\tau_0]$,
$\tau_0>0$.
Hence, we obtain 
\begin{align}
(X_N^{A,\ve})_{\cdot+\alpha}-X_N^{A,\ve}
&
=\sum_{k=1}^N\ve^k\sum_{\whj_m\in\at_k}
\left(A_{k,\whj_m}({\cdot},{\cdot}{+}\ve\alpha)\e^{\i\vth_{\whj_m}\cdot\alpha}
-A_{k,\whj_m}\right)\boe_{\whj_m}
\notag\\&
=\sum_{k=1}^N\ve^k\sum_{\whj_m\in\at_k}
\left(\sum_{s=0}^{N-k+1}\ve^sD_{k,s,\whj_m,\alpha}\right)\boe_{\whj_m}
\notag\\&
=\sum_{k=1}^{N+1}\ve^k d_{k,\alpha},\qquad 
d_{k,\alpha}:=\sum_{\whj_m\in\at_k}\sum_{\ell=m}^k 
D_{\ell,k-\ell,\whj_m,\alpha}\boe_{\whj_m},
\label{XNAalphagamma}
\end{align}
with
\begin{align}\label{D}
D_{k,s,\whj_m,\alpha}:=
\begin{cases}
\ds\left(\e^{\i\vth_{\whj_m}\cdot\alpha}{-}1\right)A_{k,\whj_m},
& s=0,
\\[3mm] 
\ds\e^{\i\vth_{\whj_m}\cdot\alpha}
\frac{(\alpha{\cdot}\nabla_y)^sA_{k,\whj_m}}{s!},
&s=1,\ldots,N{-}k,
\quad k<N,
\\[3mm]
\ds\e^{\i\vth_{\whj_m}\cdot\alpha}
\frac{(\alpha{\cdot}\nabla_y)^{N-k+1}A_{k,\whj_m}^{(h)}}{(N{-}k{+}1)!},
& s=N{-}k{+}1,
\end{cases}
\end{align}
where $A_{k,\whj_m}^{(h)}:=A_{k,\whj_m}({\cdot},{\cdot}{+}h\ve\alpha)$,
$h=h_{k,\whj_m,\ve\g}(\ve\alpha)\in(0,1)$.
%
%
(Note, that $D_{N+1,0,\whj_m,\alpha}=0$ 
and $D_{k,s,\whj_m,\alpha}=0$ for $m>k$.)
Next, abbreviating 
\begin{align}\label{XNAak}
X_N^{A,\ve}:=\sum_{k=1}^N \ve^k a_k,
\quad 
a_k:=\sum_{\whj_m\in\at_k}
A_{k,\whj_m}\boe_{\whj_m}
\end{align} 
(cf.\ \eqref{XNA}), we use \eqref{XNAalphagamma}, \eqref{XNAak}
in order to expand the first sum on the right-hand side of 
\eqref{insertedansatz} in terms of $\ve^k$,
\begin{align*}
&
\sum_{n=1}^{N}
 \left(\sum_{\alpha\in\Gamma}a_{n,\alpha}
 \left(\sum_{k=1}^{N+1}\ve^kd_{k,\alpha}\right)^n
 -b_n\left(\sum_{k=1}^N\ve^ka_k\right)^n\right)
\\ &\qquad=\sum_{n=1}^{N}
\sum_{k=n}^{n(N+1)}\ve^{k}\sum_{k_1+\ldots+k_n=k\atop 1\le k_i\le N+1}
\left(\sum_{\alpha\in\Gamma}a_{n,\alpha}\prod_{i=1}^n d_{k_i,\alpha}
-b_n\prod_{i=1}^n a_{k_i}\right)
\\ &\qquad
=\sum_{k=1}^{N(N+1)}\ve^{k}
s_k,\qquad s_k:=
\sum_{n=\left[\frac{k-1}{N+1}\right]+1}^{\min\{k,N\}}
\sum_{k_1+\ldots+k_n=k\atop 1\le k_i\le N+1}
\left(\sum_{\alpha\in\Gamma}a_{n,\alpha}\prod_{i=1}^n d_{k_i,\alpha}
   -b_n\prod_{i=1}^n a_{k_i}\right)
\end{align*}
(with $x=[x]+h$, $[x]\in\Z$, $h\in[0,1)$ for $x\in\R$).
%
It remains to obtain the expansion of $s_k$ in terms of $\boe_{\whj_m}$.
By \eqref{XNAak}, \eqref{XNAalphagamma} we get 
for the summand in brackets in $s_k$
%
\begin{align*}
&
\sum_{\alpha\in\Gamma}a_{n,\alpha}\prod_{i=1}^n
\sum_{\whj_{m_i}\in\at_{k_i}}\sum_{\ell_i=m_i}^{k_i}
D_{\ell_i,k_i-\ell_i,\whj_{m_i},\alpha}\boe_{\whj_{m_i}}
-b_n\prod_{i=1}^n
\sum_{\whj_{m_i}\in\at_{k_i}}
A_{k_i,\whj_{m_i}}\boe_{\whj_{m_i}}
\\&\qquad
=\sum_{\whj_m\in\at_k}
\sum_{(\whj_{m_1},\ldots,\whj_{m_n})=\whj_m\atop\whj_{m_i}\in\at_{k_i}}
\left(
\sum_{\ell_1+\ldots+\ell_n\le k\atop m_i\le \ell_i\le k_i}
\sum_{\alpha\in\Gamma}a_{n,\alpha}
\prod_{i=1}^n D_{\ell_i,k_i-\ell_i,\whj_{m_i},\alpha}
-b_n\prod_{i=1}^n
A_{k_i,\whj_{m_i}}\right)
\boe_{\whj_m},
\end{align*}
and hence
by inserting the last expression in the definition of $s_k$ above
\begin{multline}\label{sk}
s_k=
\sum_{\whj_m\in\at_k}
\sum_{n=\left[\frac{k-1}{N+1}\right]+1}^{\min\{k,N\}}
\sum_{k_1+\ldots+k_n=k\atop 1\le k_i\le N+1}
\sum_{(\whj_{m_1},\ldots,\whj_{m_n})=\whj_m\atop\whj_{m_i}\in\at_{k_i}}
\\
\Bigg(c_{(\whj_{m_1},\ldots,\whj_{m_n})}\prod_{i=1}^n A_{k_i,\whj_{m_i}}
+\sum_{\ell_1+\ldots+\ell_n\le k-1\atop m_i\le\ell_i\le k_i}
\sum_{\alpha\in\Gamma}a_{n,\alpha}
\prod_{i=1}^n D_{\ell_i,k_i-\ell_i,\whj_{m_i},\alpha}\Bigg)
\boe_{\whj_m}
\end{multline}
with 
\begin{align}\label{c}
c_{(\whj_{m_1},\ldots,\whj_{m_n})}:=\sum_{\alpha\in\Gamma}a_{n,\alpha}
\prod_{i=1}^n \left(\e^{\i\vth_{\whj_{m_i}}\cdot\alpha}{-}1\right)-b_n.
\end{align}
In particular for $n=2$, 
recalling $a_{2,\alpha}=-a_{2,-\alpha}$ for $\alpha\in\Gamma$,
we obtain
\begin{align}\label{c2gen}
c_{(\whj_m,\whj_\mu)}=
-4\i\sum_{\alpha\in\Gamma}a_{2,\alpha}
\sin\left(\frac{\vth_{\whj_m}}2{\cdot}\alpha\right)
\sin\left(\frac{\vth_{\whj_\mu}}2{\cdot}\alpha\right)
\sin\left(\frac{\vth_{\whj_m}{+}\vth_{\whj_\mu}}2{\cdot}\alpha\right)
-b_2.
\end{align}
For $k=1,\ldots,N$ the involved expansion \eqref{sk} can be structured by
separating the terms involving the functions $A_{k,\whj_m}$ and 
$A_{k-1,\whj_m}$ from all the others, which we subsum into the term
$\calS_{k-2,\whj_m}$. This structure is 
essential 
for the derivation of the macroscopic equations for $A_{k,\whj_m}$.
Indeed, by writing out the summands for $n=1$ and $n=2$ in \eqref{sk} 
when $k=1,\ldots,N$, we obtain
\begin{multline}\label{skN}
s_k=
\sum_{\whj_m\in\at_k}
\Big(
c_{(\whj_{m})}A_{k,\whj_{m}}
+
2\i\Om(\vth_{\whj_m})\nabla_\vth\Om(\vth_{\whj_m}){\cdot}\nabla_y 
A_{k-1,\whj_m}
\\
+
\min\{k{-}1,2\}
\sum_{(p,\whj_{\mu})=\whj_m
\atop p\in\calN,\whj_{\mu}\in\at_{k-1}}
c_{(p,\whj_{\mu})}A_{1,p} A_{k-1,\whj_{\mu}}
+\calS_{k-2,\whj_m}\Big)\boe_{\whj_m}
\end{multline}
with
\begin{align}
\calS_{k-2,\whj_m}:=
&
\sum_{\ell=m}^{k-2}
\sum_{\alpha\in\Gamma}a_{1,\alpha}D_{\ell,k-\ell,\whj_{m},\alpha}
+\sum_{\kappa=2}^{k-2}
\sum_{(\whj_{m_1},\whj_{m_2})=\whj_m
\atop\whj_{m_1}\in\at_{\kappa},\whj_{m_2}\in\at_{k-\kappa}}
c_{(\whj_{m_1},\whj_{m_2})}
A_{\kappa,\whj_{m_1}} A_{k-\kappa,\whj_{m_2}}
\notag\\
&+
\sum_{k_1+k_2=k\atop 1\le k_i\le k-1}
\sum_{(\whj_{m_1},\whj_{m_2})=\whj_m\atop\whj_{m_i}\in\at_{k_i}}
\sum_{\ell_1+\ell_2\le k-1\atop m_i\le \ell_i\le k_i}
\sum_{\alpha\in\Gamma}a_{2,\alpha}
\prod_{i=1}^2 D_{\ell_i,k_i-\ell_i,\whj_{m_i},\alpha}
\notag\\&
+\sum_{n=3}^{k}
\sum_{k_1+\ldots+k_n=k\atop 
1\le k_i\le k-n+1}
\sum_{(\whj_{m_1},\ldots,\whj_{m_n})=\whj_m\atop\whj_{m_i}\in\at_{k_i}}
\notag\\&\qquad
\Bigg(
c_{(\whj_{m_1},\ldots,\whj_{m_n})}
\prod_{i=1}^n A_{k_i,\whj_{m_i}}
+\sum_{\ell_1+\ldots+\ell_n\le k-1\atop m_i\le \ell_i\le k_i}
\sum_{\alpha\in\Gamma}a_{n,\alpha}
\prod_{i=1}^n D_{\ell_i,k_i-\ell_i,\whj_{m_i},\alpha}
\Bigg).
\label{source}
\end{align}
For \eqref{skN} we used
\begin{equation}\label{spacederivative}
\sum_{\alpha\in\Gamma}a_{1,\alpha}D_{k-1,1,\whj_{m},\alpha}
=\sum_{\alpha\in\Gamma}a_{1,\alpha}\e^{\i\vth_{\whj_m}\cdot\alpha}
\alpha{\cdot}\nabla_y A_{k-1,\whj_m}
=2\i\Om(\vth_{\whj_m})\nabla_\vth\Om(\vth_{\whj_m}){\cdot}\nabla_y 
A_{k-1,\whj_m}
\end{equation}
(cf.\ \eqref{D} and \eqref{DR}) for the term obtained for $n=1$, 
and the formula 
\begin{equation*}\sum_{\kappa=1}^{k-1}a_{\kappa,k-\kappa}
=\min\{k{-}1,2\}a_{1,k-1}+\sum_{\kappa=2}^{k-2}a_{\kappa,k-\kappa}
\quad\text{for}\quad a_{\kappa,k-\kappa}=a_{k-\kappa,\kappa}
\end{equation*}
for the term obtained for $n=2$. 
(Note, that $\calS_{k-2,\whj_m}$ contains indeed 
only functions $A_{\ell,\whj_m}$ with $\ell\le k-2$.)
Finally, since $V^\prime_{N+1,\alpha}(x)=\OO(|x|^{N+1})$,
$W^\prime_{N+1}(x)=\OO(|x|^{N+1})$ (cf.\ \eqref{VW})
and $(X_N^{A,\ve})_{\cdot+\alpha}{-}X_N^{A,\ve}=\OO(\ve)$,
$X_N^{A,\ve}=\OO(\ve)$ 
pointwise for $(t,\gamma)\in[0,\infty)\times\Gamma$
(cf.\ \eqref{XNAalphagamma}, \eqref{XNA}),
the second sum on the right-hand side of \eqref{insertedansatz} is of order 
$\OO(\ve^{N+1})$.\\

Using the above expansions, 
\eqref{insertedansatz} reads
\begin{equation}\label{onwards}
\sum_{k=1}^{N}\ve^{k}(t_k-s_k)+\res\big(X_N^{A,\ve}\big)=0
\end{equation}
with $t_k$, $s_k$ given by \eqref{tk}, \eqref{skN}, 
and the {\it residuum} 
$\res\big(X_N^{A,\ve}\big)=\OO(\ve^{N+1})$
given by 
 \begin{align}\label{res}
 \res\big(X_N^{A,\ve}\big)
 :=&\,\ve^{N+1}t_{N+1}+\ve^{N+2}t_{N+2}
 -\sum_{k=N+1}^{N(N+1)}\ve^k s_k
 \notag\\&-\sum_{\alpha\in\Gamma}V^\prime_{N+1,\alpha}
 \big((X_N^{A,\ve})_{\cdot+\alpha}{-}X_N^{A,\ve}\big)
 +W^\prime_{N+1}\big(X_N^{A,\ve}\big)
\end{align}
with $s_k$ given by \eqref{sk}.
Thus, the approximation 
$X_N^{A,\ve}$ 
\emph{satisfies the lattice system 
\eqref{Mexpl} 
pointwise for $(t,\gamma)\in[0,\infty)\times\Gamma$
up to order $\ve^N$ 
if and only if} 
\begin{equation}\label{onwards2}
t_k-s_k=0\quad\text{for all}\quad k=1,\ldots,N.
\end{equation}
Furthermore, by \eqref{tk} and \eqref{skN}, 
$t_k-s_k$ is given for each $k=1,\ldots,N$ 
as an expansion in terms of the 
harmonic functions $\boe_{\whj_m}$, $\whj_m\in\at_k$. 
%
Since the coefficients of this expansion, 
the macroscopic amplitudes $A_{k,\whj_m}$ and their derivatives,
are varying much slower in space and time 
as compared to the microscopically oscillating, mutually different, 
non-vanishing patterns $\boe_{\whj_m}$ 
(with different wave-vectors and frequencies and modulus $1$),
in order for the expansion $t_k-s_k$ to equal $0$, 
each of its coefficients has to vanish identically.
Hence, the approximation 
$X_N^{A,\ve}$ satisfies 
the lattice system 
\eqref{Mexpl} up to order $\ve^N$ if and only if
\begin{align}
\delta_{\whj_m} A_{k,\whj_m}
+2\i\om_{\whj_m}\partial_\tau A_{k-1,\whj_m}
-2\i\Om(\vth_{\whj_m})\nabla_\vth\Om(\vth_{\whj_m}){\cdot}\nabla_y 
A_{k-1,\whj_m}&
\notag\\
-\min\{k{-}1,2\}\sum_{(p,\whj_{\mu})=\whj_m
\atop p\in\calN,\whj_{\mu}\in\at_{k-1}}
c_{(p,\whj_{\mu})} 
A_{1,p} A_{k-1,\whj_{\mu}}
&=\calS_{k-2,\whj_m}-\partial_\tau^2 A_{k-2,\whj_m}
\notag\\
\text{for all $k=1,\ldots,N$}&\text{ and all $\whj_m\in\at_k$,}
\label{tk-sk}
\end{align}
where we used $-\om_{\whj_m}^2-c_{(\whj_m)}
=-\om_{\whj_m}^2+\Om^2(\vth_{\whj_m})=\delta_{\whj_m}$, 
cf.\  \eqref{c}, \eqref{DR} and \eqref{closedness}.
%
%
The equations \eqref{tk-sk} are the \emph{macroscopic equations}, 
which the functions $A_{k,\whj_m}$
of the approximation $X_N^{A,\ve}$ given by \eqref{XNA}
have to satisfy necessarily,
in order for this approximation to solve the microscopic model
\eqref{Mexpl} 
up to residual terms of order $\ve^{N+1}$.
However, the equation for $A_{k,-\whj_m}$ is just the complex conjugate 
of the equation for $A_{k,\whj_m}$. 
Thus, since 
$A_{k,-\whj_m}=\ol{A_{k,\whj_m}}$, 
it suffices to determine just one of them.

Since the equations for any given $k=1,\ldots,N$ involve the functions  
$A_{\kappa,\whj_m}$ with $\kappa\le k$ 
it is natural that they have to be solved inductively for increasing $k$.
Hence, recalling that $A_{k,\whj_m}=0$ for $k\in\Z\setminus\{1,\ldots,N\}$ 
(and thus in particular $\calS_{k-2,\whj_m}=0$ for $k\le 2$) 
and $m>k$,
the equations \eqref{tk-sk} for $k=1$ read
\begin{equation}\label{macro0}
\delta_j A_{1,j}=0\qquad\text{for $j\in\calN$.}
\end{equation}
Since by assumption $\delta_j=\Om^2(\vth_j)-\om_j^2=0$, reflecting the fact
that the first-order approximation $X_1^{A,\ve}$ in \eqref{XNA} consists of
modulated pulses, these equations are automatically fulfilled 
and the amplitudes $A_{1,j}$ remain at this stage undetermined.
%
Of course, 
these equations can be interpreted also in the opposite direction, 
namely as requiring necessarily from the pairs $\pm(\vth_j,\om_j)$ to satisfy 
the dispersion relations 
$\delta_j=0$, i.e.\ to characterize pulses, 
in order to allow for non-vanishing amplitudes $A_{1,j}$ 
in the approximation $X_1^{A,\ve}$.
%
%
%
Then, 
the functions $A_{1,j}$ are determined by the equations \eqref{tk-sk} 
for $k=2$ with $\delta_j=0$
\begin{align}\label{macro1}
2\i\om_j\partial_\tau A_{1,j}
-2\i\Omega(\vth_j)\nabla_\vth\Omega(\vth_j){\cdot}\nabla_y A_{1,j}
=\sum_{(p,q)=j\atop p,q\in\calN}c_{(p,q)}A_{1,p}A_{1,q}
\qquad\text{for $j\in\calN$}
\end{align}
and
\begin{align}\label{macro2}
\delta_{\whj_2} A_{2,\whj_2}
=\sum_{(p,q)=\whj_2\atop p,q\in\calN}c_{(p,q)}A_{1,p}A_{1,q}
\qquad\text{for $\whj_2\in\calN^2\setminus\calN$,}
\end{align}
where 
by \eqref{c2gen}
\begin{align}
c_{(p,q)}&=
-4\i\sum_{\alpha\in\Gamma}a_{2,\alpha}
\sin\left(\frac{\vth_p}2{\cdot}\alpha\right)
\sin\left(\frac{\vth_q}2{\cdot}\alpha\right)
\sin\left(\frac{\vth_p{+}\vth_q}2{\cdot}\alpha\right)
-b_2.
\label{c2}
\end{align}
%
%
%
The equations \eqref{macro1} determine the amplitudes $A_{1,j}$, $j\in\calN$.
%
If for some $j\in\calN$ there exist $p,q\in\calN$ such that $(p,q)=j$
then the corresponding equations are semilinearly coupled transport equations. 
If for some $j\in\calN$ there do not exist such $p,q\in\calN$, then the
corresponding equation for $A_{1,j}$ is just a transport equation with
vanishing right hand side, uncoupled from the other equations.
The same applies if the coupling coefficient 
$c_{(p,q)}$ given by 
\eqref{c2} vanishes, for instance when the interaction and on-site potentials 
$V_\alpha$ and $W$, cf.\ \eqref{VW}, have no cubic terms. 
%
%
Of course, within the considered system of pulses $\calN$ 
there can exist subsystems of pulses interacting with each other but
not interacting with other (subsystems of) pulses.
The corresponding macroscopic equations 
then establish (closed) coupled subsystems, cf.\ also Sec.\ \ref{secexamples}.
Note that each amplitude $A_{1,j}$ is transported by the group velocity
$\nabla_\vth\Omega(\vth_j)$ of its corresponding pulse.

Having determined the amplitudes $A_{1,j}$ for $j\in\calN$ 
by \eqref{macro1}, 
we are then able to calculate the functions
$A_{2,\whj_2}$ for $\whj_2\in\calN^2\setminus\calN$ 
by 
\eqref{macro2}, 
provided $\delta_{\whj_2}\neq 0$.
If $\delta_{\whj_2}=0$, then 
$\pm(\vth_{\whj_2},\om_{\whj_2})=\pm(\vth_p+\vth_q,\om_p+\om_q)$ 
characterizes a pulse not considered 
(or, equivalently, assumed to have an identically vanishing amplitude) 
in our approximation $X_1^{A,\ve}$,
and \eqref{macro2} can be seen as a further condition on 
the first order amplitudes $A_{1,p}$, $A_{1,q}$ for which $(p,q)=\whj_2$.
As we will exemplify in Section \ref{secexamples}, 
this condition then implies that 
some of the involved first order amplitudes 
have also to vanish identically,
thus 
effectively 
prohibiting the description of the macroscopic dynamics of the 
corresponding pulses.
This problem can be overcome 
if we include into $\calN$ all $(p,q)\in\calN^2$ with $\delta_{(p,q)}=0$.
Then $\delta_{\whj_2}\neq0$ for $\whj_2\in\calN^2\setminus\calN$,
and we call such a set of pulses 
$\{\pm(\vth_j,\om_j):\ j=1,\ldots,\nu\}$
\emph{closed under interactions up to order $k=2$},
according to Definition \ref{defclosure}. 
%
Then, 
for a set $\calN$ of pulses with this property, 
we can calculate by \eqref{tk-sk} for $k=2$ the functions   
$A_{2,\whj_2}$ for $\whj_2\in\calN^2\setminus\calN$,
while the second order amplitudes $A_{2,j}$, $j\in\calN$, remain 
undetermined by these equations.

%
%
%
%


The appearing pattern, namely that the functions 
$A_{k,\whj_m}$ with $\whj_m\in\at_k\setminus\calN$
can be determined by the equations \eqref{tk-sk} for $k$ 
and the $k$-th order amplitudes $A_{k,j}$ with $j\in\calN$ 
by the equations 
for $k+1$, can be continued inductively,
provided  the set of pulses $\calN$ is closed under interactions up to order 
$k$. 
More precisely, for $k=3,\ldots,N$, 
and using $\delta_j=0$ for $j\in\calN$,
the equations \eqref{tk-sk} read
\begin{multline}\label{macro3}
2\i\om_j\partial_\tau A_{k-1,j}
-2\i\Omega(\vth_j)\nabla_\vth\Omega(\vth_j){\cdot}\nabla_y A_{k-1,j}
-2\sum_{(p,q)=j\atop p,q\in\calN}c_{(p,q)}A_{1,p}A_{k-1,q}
\\
=2\sum_{(p,\whj_\mu)=j
\atop p\in\calN,\whj_\mu\in\at_{k-1}\setminus\calN}
c_{(p,\whj_\mu)}A_{1,p}A_{k-1,\whj_\mu}
+\calS_{k-2,j}
-\partial_\tau^2 A_{k-2,j}
\qquad\text{for $j\in\calN$}
\end{multline}
and 
\begin{multline}\label{macro4}
\delta_{\whj_m} A_{k,\whj_m}
=
-2\i\om_{\whj_m}\partial_\tau A_{k-1,\whj_m}
+2\i\Omega(\vth_{\whj_m})\nabla_\vth\Omega(\vth_{\whj_m})
{\cdot}\nabla_y A_{k-1,\whj_m}
\\
+2\sum_{(p,\whj_\mu)=\whj_m
\atop p\in\calN,\whj_\mu\in\at_{k-1}}
c_{(p,\whj_\mu)}A_{1,p}A_{k-1,\whj_\mu}
+\calS_{k-2,\whj_m}
-\partial_\tau^2 A_{k-2,\whj_m}
\qquad\text{for $\whj_m\in
\at_k\setminus\calN$.}
\end{multline}
The equations \eqref{macro3} form a system of 
(in general) 
linearly coupled inhomogeneous linear transport equations 
for the $(k-1)$-th order amplitudes $A_{k-1,j}$, $j\in\calN$, travelling
again with the group velocity of the pulse they modulate.
As in the case for $k=2$, the equations 
for $j\in\calN$ for which no $p,q\in\calN$ with $(p,q)=j$ exist 
decouple from the system.
Note, that the source term on the right hand side of \eqref{macro3}
is known, since it consists of functions $A_{\kappa,\whj_m}$ with 
either $\kappa=1,\ldots,k-2$ 
or $\kappa=k-1$ and $\whj_m\in\at_{k-1}\setminus\calN$,
which have been determined 
by the previous equations \eqref{tk-sk} for $2,\ldots,k-1$.

Then, since  
by \eqref{macro3}
now
also the amplitudes $A_{k-1,j}$ with $j\in\calN$ 
are determined, 
the right hand sides of the equations \eqref{macro4} are known,
and we can determine by these equations the functions $A_{k,\whj_m}$
for $\whj_m\in\at_k\setminus\calN$,
provided the system $\calN$ of the pulses under consideration
is closed under interactions up to order $k$, 
which guarantees $\delta_{\whj_m}\neq0$.
The $k$-th order amplitudes $A_{k,j}$, $j\in\calN$, remain undetermined.
Hence, performing the above procedure inductively up to $k=N$,
all equations \eqref{tk-sk} or, equivalently, \eqref{onwards2}
are satisfied, while no conditions are imposed on $A_{N,j}$.
Thus, we obtain the following result, 
which establishes the formal derivation of the macroscopic dynamics.
%
\begin{theorem}\label{formalderivation}
Let $\{\pm(\vth_j,\om_j): j=1,\ldots,\nu\}$ be a set of $\nu\in\N$ 
different pulses, closed under interactions up to order $N$ according to
Definition \ref{defclosure}, 
and let the amplitudes  
$A_{k,j}$ 
of the multiscale ansatz $X_N^{A,\ve}$ given by \eqref{XNA}
satisfy the macroscopic equations 
\eqref{macro1} for $k=1$, \eqref{macro2} for $k=2,\ldots,N{-}1$, 
and set $A_{N,j}=0$.
Moreover, set the functions $A_{k,\whj_m}$, $\whj_m\in\at_k\setminus\calN$, 
according to \eqref{macro3} for $k=2$, \eqref{macro4} for $k=3,\ldots,N$.
%
Then, the ansatz $X_N^{A,\ve}$ constructed by these functions 
satisfies the microscopic model \eqref{M} up to order $\ve^N$, i.e.
for the residual terms given by \eqref{res} we have 
$\res\big(X_N^{A,\ve}\big)=\OO\big(\ve^{N+1}\big)$.
\end{theorem}
%
%

However, all results obtained in this section are formal
in the sense that they follow from the a priori assumption that solutions to
\eqref{M} {\it retaining over time} the form $x=X_{N}^{A,\ve}+\OO(\ve^{N+1})$
exist.
%
%
Whether this is indeed the case, i.e. whether 
an approximation $X_{N}^{A,\ve}$ constructed by the solutions of the
derived macroscopic equations {\it stays close} 
to an original solution of \eqref{M} 
{\it over macroscopic time intervalls},
is discussed in the next section.
\section{Justification}\label{jf}
%
\subsection{Estimate of the residuum}
In order to justify the macroscopic equations derived in the previous section,
we will need an estimate of the residual terms 
$\res\big(X_N^{A,\ve}\big)$, see \eqref{res}, 
with respect to the $\ell^2(\Gamma)$-norm,
$\ds\|x\|_{\ell^2}^2=\sum_{\g\in\Gamma}|x_\g|^2$.
%
%
%
From \eqref{res} we obtain
 \begin{align*}
  \big\|\res\big(X_N^{A,\ve}\big)\big\|_{\ell^2}
\le &\, \ve^{N+1}\|t_{N+1}\|_{\ell^2}
+\ve^{N+2}\|t_{N+2}\|_{\ell^2}
  + \sum_{k=N+1}^{N(N+1)}\ve^k\|s_k\|_{\ell^2}
\\  &\,+\sum_{\alpha\in\Gamma}\big\|V^\prime_{N+1,\alpha}
  \big(\big(X_N^{A,\ve}\big)_{\cdot+\alpha}{-}X_N^{A,\ve}
\big)\big\|_{\ell^2}
  +\big\|W^\prime_{N+1}\big(X_N^{A,\ve}\big)\big\|_{\ell^2},
 \end{align*}
provided of course the series over $\alpha\in\Gamma$ exists.
From the definition of $t_{k}$ in \eqref{tk} we obtain 
\begin{align}\label{esttN+1}
\|t_{N+1}\|_{\ell^2}
+\ve\|t_{N+2}\|_{\ell^2}
\le
\sum_{\whj_m\in\at_{N}}
\big(2|\om_{\whj_m}|\,\|\partial_\tau A_{N,\whj_m}\|_{\ell^2}
+\ve_0\|\partial_\tau^2 A_{N,\whj_m}\|_{\ell^2}
+\|\partial_\tau^2 A_{N-1,\whj_m}\|_{\ell^2}\big)
\end{align}
for $\ve\in[0,\ve_0]$, $\ve_0>0$ (with $A_{N-1,\whj_m}=0$ when $N=1$ or $m=N$).
For $s_k$ given by \eqref{sk} we first estimate (cf.\ \eqref{D}) 
\begin{equation}\label{estD}
|D_{\ell,k-\ell,\whj_{m},\alpha}|
\le 2|(\alpha{\cdot}\nabla_y)^{k-\ell}A_{\ell,\whj_m}|
\le 2|\alpha|^{k-\ell}\sum_{|\sigma|=k-\ell}
|D_y^\sigma A_{\ell,\whj_m}|
\end{equation}
for $\ell=1,\ldots,N,\ k=\ell,\ldots,N{+}1$ 
(with $A_{\ell,\whj_m}=A_{\ell,\whj_m}^{(h)}$ when $k=N{+}1$), 
where we used
  \begin{align*}
 |(\alpha{\cdot}\nabla_y)^sA_{k,\whj_m}|
   &=
\Big|\Big(\prod_{i=1}^s\sum_{d_i=1}^d\alpha_{d_i}\partial_{y_{d_i}}
  \Big)
   A_{k,\whj_m}\Big|
   =
\Big|\sum_{d_1,\ldots,d_s=1}^d
   \Big(\prod_{i=1}^s\alpha_{d_i}\partial_{y_{d_i}}\Big)A_{k,\whj_m}
   \Big|
   \\
   &
  \le\sum_{d_1,\ldots,d_s=1}^d\Big(\prod_{i=1}^s|\alpha_{d_i}|\Big)\,
   \Big|\Big(\prod_{i=1}^s\partial_{y_{d_i}}\Big)A_{k,\whj_m}\Big|
  \le
  |\alpha|^s\sum_{|\sigma|=s}|D_y^\sigma A_{k,\whj_m}|
  \end{align*}
  with $\alpha=(\alpha_1,\ldots,\alpha_d)\in\R^d$
%
and $D_y^\sigma=\partial_{y_1}^{\sigma_1}\cdots\partial_{y_d}^{\sigma_d}$,
$|\sigma|=\sigma_1{+}\ldots{+}\sigma_d$.
With \eqref{estD} and \eqref{c} we obtain
 \begin{align*}
&\Big\|c_{(\whj_{m_1},\ldots,\whj_{m_n})}\prod_{i=1}^n A_{k_i,\whj_{m_i}}
 +\sum_{\ell_1+\ldots+\ell_n\le k-1\atop m_i\le\ell_i\le k_i}
 \sum_{\alpha\in\Gamma}a_{n,\alpha}
 \prod_{i=1}^n D_{\ell_i,k_i-\ell_i,\whj_{m_i},\alpha}\Big\|_{\ell^2}
\\&\quad
\le
c_{a,b,k,n}
\sum_{\ell_1+\ldots+\ell_n\le k\atop m_i\le\ell_i\le k_i}
\Big(\prod_{i=1}^{n-1}\sum_{|\sigma|=k_i-\ell_i}
\|D_y^\sigma A_{\ell_i,\whj_{m_i}}\|_{\ell^\infty}
\Big)
\sum_{|\sigma|=k_n-\ell_n}
\|D_y^\sigma A_{\ell_n,\whj_{m_n}}\|_{\ell^2}
 \end{align*}
with $\ds \|x\|_{\ell^\infty}=\max_{\g\in\Gamma}|x_\g|$,
$\ds c_{a,b,k,n}:=2^n\sum_{\alpha\in\Gamma}|a_{n,\alpha}|
(1{+}|\alpha|^{k-n})+|b_n|$,
and where we used that $(n\le)\ell_1+\ldots+\ell_n=k=k_1+\ldots+k_n$ 
with $\ell_i\le k_i$ implies $\ell_i=k_i$.
Hence, by \eqref{sk} we get
\begin{multline}\label{estsN+1}
\sum_{k=N+1}^{N(N+1)}\ve^{k} \left\|s_k\right\|_{\ell^2}
\le
\ve^{N+1}
c_{a,b}
\sum_{k=N+1}^{N(N+1)}\ve_0^{k-N-1}
\sum_{n=\left[\frac{k-1}{N+1}\right]+1}^{N}
\sum_{k_1+\ldots+k_n=k\atop 1\le k_i\le N+1}
\sum_{(\whj_{m_1},\ldots,\whj_{m_n})\in\at_{k}
\atop\whj_{m_i}\in\at_{k_i}}
\sum_{\ell_1+\ldots+\ell_n\le k\atop m_i\le\ell_i\le k_i}
\\
\Big(\prod_{i=1}^{n-1}\sum_{|\sigma|=k_i-\ell_i}
\|D_y^\sigma A_{\ell_i,\whj_{m_i}}\|_{\ell^\infty}\Big)
\sum_{|\sigma|=k_n-\ell_n}
\|D_y^\sigma A_{\ell_n,\whj_{m_n}}\|_{\ell^2}
\end{multline}
for $\ve\in[0,\ve_0]$ 
with 
$\ds c_{a,b}:=\max_{n=1,\ldots,N}\Big(2^{n}\sum_{\alpha\in\Gamma}
|a_{n,\alpha}|(2+|\alpha|^{nN})+|b_n|\Big)$
and $A_{\ell_i,\whj_{m_i}}=A_{\ell_i,\whj_{m_i}}^{(h)}$ when $k_i=N{+}1$. 
Finally, since by $V_\alpha,\ W\in\Ce^{N+2}(\R)$ 
\begin{equation}\label{VWtilde}
\big|V^\prime_{N+1,\alpha}(x)\big|\le a_{N+1,\alpha,x_0}|x|^{N+1},
\qquad\big|W^\prime_{N+1}(x)\big|\le b_{N+1,x_0}|x|^{N+1}
\end{equation}
for $x\in[-2x_0,2x_0]$, $x_0>0$,
with $a_{N+1,\alpha,x_0},\ b_{N+1,x_0}>0$
(cf.\ \eqref{VW}),
and since
\begin{align}\label{estX}
\big\|X_N^{A,\ve}\big\|_{\ell^\infty}
\le\ve\sum_{k=1}^N \ve_0^{k-1}
\sum_{\whj_m\in\at_k}
\|A_{k,\whj_m}\|_{\ell^\infty},
\qquad
\big\|X_N^{A,\ve}\big\|_{\ell^2}
\le\ve\sum_{k=1}^N \ve_0^{k-1}
\sum_{\whj_m\in\at_k}
\|A_{k,\whj_m}\|_{\ell^2}
\end{align}
(cf.\ \eqref{XNA}) for $\ve\in[0,\ve_0]$, 
we obtain, considering 
$\big\|\big(X_N^{A,\ve}\big)_{\cdot+\alpha}\big\|_{\ell^2}
=\big\|X_N^{A,\ve}\big\|_{\ell^2}$ for $\alpha\in\Gamma$,
\begin{align}
&
\sum_{\alpha\in\Gamma}\big\|V^\prime_{N+1,\alpha}
  \big(\big(X_N^{A,\ve}\big)_{\cdot+\alpha}{-}X_N^{A,\ve}\big)
  \big\|_{\ell^2}
  +\big\|W^\prime_{N+1}\big(X_N^{A,\ve}\big)\big\|_{\ell^2}
\notag\\&
  \le\sum_{\alpha\in\Gamma}
  a_{N+1,\alpha,x_0}
  \big\|\big(X_N^{A,\ve}\big)_{\cdot+\alpha}{-}X_N^{A,\ve}
  \big\|_{\ell^\infty}^{N}
  \big\|\big(X_N^{A,\ve}\big)_{\cdot+\alpha}{-}X_N^{A,\ve}\big\|_{\ell^2}
  +b_{N+1,x_0}
  \big\|X_N^{A,\ve}\big\|_{\ell^\infty}^N\big\|X_N^{A,\ve}\big\|_{\ell^2}
%
%
\notag\\&
\le\ve^{N+1}
c_{V,W,x_0}
\Big(\sum_{k=1}^N\ve_0^{k-1}\sum_{\whj_m\in\at_k}
\|A_{k,\whj_m}\|_{\ell^\infty}\Big)^{N} 
\sum_{k=1}^N\ve_0^{k-1}\sum_{\whj_m\in\at_k}
\|A_{k,\whj_m}\|_{\ell^2}
\label{estVWtilde}
\end{align}
for $\ve\in[0,\ve_0]$ and 
$\ds 
\sum_{k=1}^N \ve_0^{k}\sum_{\whj_m\in\at_k}\|A_{k,\whj_m}\|_{\ell^\infty}
\le x_0$
with $\ds c_{V,W,x_0}
:=2^{N+1}\sum_{\alpha\in\Gamma}a_{N+1,\alpha,x_0}+b_{N+1,x_0}$.\\
Hence, for potentials $V_\alpha, W\in\Ce^{N+2}(\R)$ ($\alpha\in\Gamma$), 
cf.\ \eqref{VW}, which satisfy
\begin{align}
&\max_{n=1,\ldots,N}
\Big(\sum_{\alpha\in\Gamma}|a_{n,\alpha}||\alpha|^{nN}\Big)<\infty
\quad\text{and}
\notag\\&
\eqref{VWtilde},\qquad
\sum_{\alpha\in\Gamma}a_{N+1,\alpha,x_0}<\infty
\quad\text{with}\quad
x_0\ge \sum_{k=1}^N \ve_0^{k}\sum_{\whj_m\in\at_k}
\|A_{k,\whj_m}\|_{\ell^\infty},
\label{condalpha}
\end{align}
we obtain by \eqref{esttN+1}, \eqref{estsN+1} and \eqref{estVWtilde}
for 
$\ve_0,\tau_0>0$ 
the estimate 
\begin{align}\label{estreswt}
\big\|\res\big(X_N^{A,\ve}\big)\big\|_{\ell^2}
\le \ve^{N+1} \wt{C}_r
\quad\text{for $\ve\in(0,\ve_0]$, $t\in[0,\tau_0/\ve]$,}
\end{align}
with $\wt{C}_r>0$ independent of $\ve$ and $t$, 
provided the estimates 
\begin{align}
&
\|\partial_\tau A_{N,\whj_m}\|_{\ell^2},\ 
\|\partial_\tau^2 A_{N,\whj_m}\|_{\ell^2},\
\|\partial_\tau^2 A_{N-1,\whj_m}\|_{\ell^2}
<\infty,
\notag\\&
\|D_y^\sigma A_{k,\whj_m}\|_{\ell^\infty},\
\|D_y^\sigma A_{k,\whj_m}\|_{\ell^2}
<\infty
\quad\text{for $|\sigma|=\ell-k$, $\ell=1,\ldots,N+1$, $k=1,\ldots,\ell$}
\label{discrresnorms}
\end{align}
are satisfied uniformly in $\tau=\ve t\in[0,\tau_0]$
(with $A_{k,\whj_m}=A_{k,\whj_m}^{(h)}$ when $\ell=N+1$, cf.\ \eqref{D}). 

However, the $k$-th order amplitudes 
$A_{k,j}$, $k=1,\ldots,N-1$, $j=1,\ldots,\nu$, 
are 
obtained as 
solutions of the partial
differential equations \eqref{macro1}, \eqref{macro3},
and the functions $A_{k,\whj_m}$, $k=2,\ldots,N$, 
$\whj_m\in\at_k\setminus\calN$,
are given 
via the formulas 
\eqref{macro2}, \eqref{macro4}.
All these functions depend on the (continuous) 
macroscopic time and space variables 
$\tau=\ve t$ and $y=\ve\eta$ with $\eta=\g$ for $\g\in\Gamma$. 
Thus, we have to relate the above 
$\ell^\infty(\Gamma)$- and $\ell^2(\Gamma)$-norms
to the norms of the solution spaces of the appearing 
partial differential equations. 
This is the purpose of the following lemma.
\begin{lemma}\label{DCNT}
For $d\in\N$, $s>d/2$
there exists a $c_p>0$, 
such that  
\begin{align*}
\left\|\phi\left(\ve(\cdot+h\alpha)\right)\right\|_{\ell^2}
\le c_p\, \ve^{-d/2}\|\phi\|
_{\Ha^s(\R^d)}
\end{align*}
for all $\phi\in\Ha^s(\R^d)$, 
$\alpha\in\Gamma$, 
%
%
$h\in[0,1]$ 
and all $\ve\in(0,\ve_0]$, $\ve_0>0$.
\end{lemma}
{\bf Proof:}  
By Sobolev's imbedding theorem,
%
%
there exists a $c_s>0$, 
such that 
$\|\varphi\|
_{\Le^\infty(\calC)}
\le c_s\|\varphi\|
_{\Ha^s(\calC)}
$ 
for all $\varphi\in\Ha^s(\calC)$ with
$\calC:=\{h_1g_1+\ldots+h_dg_d:\ (h_1,\ldots,h_d)\in[0,1)^d\}$ 
(cf.\ \eqref{Gamma}).
Hence, since $h\alpha=h(\alpha_1g_1+\ldots+\alpha_dg_d)
=[h\alpha_1]g_1+\ldots+[h\alpha_d]g_d+h_\ast$
%
with $\alpha_i\in\Z$, $h_\ast\in\calC$,
we get
\begin{align*}
\left\|\phi\left(\ve(\cdot+h\alpha)\right)\right\|_{\ell^2}^2
&=\sum_{\gamma\in\Gamma}\left|\phi\left(\ve(\gamma+h\alpha)\right)\right|^2
=\sum_{\gamma\in\Gamma}\left|\phi\left(\ve(\gamma+h_\ast)\right)\right|^2
\le\sum_{\gamma\in\Gamma}\left\|\phi\left(\ve(\gamma+\cdot)\right)
\right\|^2
_{\Le^\infty(\calC)}
\\&
\le c_s^2\sum_{\gamma\in\Gamma}\left\|\phi\left(\ve(\gamma+\cdot)\right)
\right\|^2
_{\Ha^s(\calC)}
=c_s^2\left\|\phi\left(\ve\cdot\right)\right\|^2
_{\Ha^s(\R^d)}
\le c^2\ve^{-d}\left\|\phi\right\|^2
_{\Ha^s(\R^d)}
\end{align*}
with $c_p:=c_s\max\{1,\ve_0^{s}\}$, obtaining the latter inequality by
rescaling.$\hfill\square$\\

\noindent Hence, under  
\begin{assumption}\label{regularityassumption}
For $d\in\N$ and $s>d/2$ there exists a $\tau_0>0$ such that 
the functions $ A_{k,\whj_m}:[0,\tau_0]\times\R^d\to\C$, 
$k=1,\ldots,N$, $\whj_m\in\at_k$,
given by \eqref{macro1}, \eqref{macro2}, \eqref{macro3}, \eqref{macro4}
satisfy
\begin{equation*}
\partial_\tau A_{N,\whj_m},\ 
\partial_\tau^2 A_{N,\whj_m},\
\partial_\tau^2 A_{N-1,\whj_m},\
D_y^\sigma A_{k,\whj_m}
\in\Ce\left([0,\tau_0];\Ha^s(\R^d;\C)\right)
\quad\text{for $|\sigma|\le N+1-k$.}
\end{equation*}
\end{assumption}
%
%
%
we obtain 
\begin{lemma}\label{lemmares}
Under 
Assumption \ref{regularityassumption}, 
and for $\ve_0>0$, $V_\alpha,W\in\Ce^{N+2}(\R)$ $(\alpha\in\Gamma)$ 
as in \eqref{VW}, \eqref{condalpha} with 
$x_0\ge \sum\limits_{k=1}^N \ve_0^{k}\sum\limits_{\whj_m\in\at_k}
\max\limits_{\tau\in[0,\tau_0]}\|A_{k,\whj_m}\|_\infty$,
there exists a $C_r>0$ independent of $\ve$ and $t$, such that
\begin{align}\label{estres}
\big\|\res\big(X_N^{A,\ve}\big)\big\|_{\ell^2}
\le \ve^{N+1-d/2} C_r
\quad\text{for $\ve\in(0,\ve_0]$, $t\in[0,\tau_0/\ve]$.}
\end{align}
\end{lemma}
{\bf Proof:} 
By Assumption \ref{regularityassumption}, Sobolev's imbedding theorem
on $\R^d$ with $s>d/2$, and Lemma \ref{DCNT}, all norms in 
\eqref{discrresnorms} are uniformly bounded for $\tau\in[0,\tau_0]$, 
and \eqref{estreswt} gives the assertion of the lemma.\hfill$\square$
%
%
\begin{remark}\label{rem_lemmares}{\em  
{\bf (a)}
Note that the order of $\ve$ asserted in \eqref{estres}  
could only be obtained due to
the linear dependence of the estimates \eqref{estsN+1}, \eqref{estVWtilde}
on the $\ell^2(\Gamma)$-norms of $D_y^\sigma A_{k,\whj_m}$.
%
Moreover, Assumption \ref{regularityassumption} 
implies by Sobolev's imbedding theorem 
the uniform boundedness in time of these functions
with respect to the $\Le^\infty(\R^d)$- 
and hence the $\ell^\infty(\Gamma)$-norm.
Furthermore, assumption and theorem yield
$A_{k,\whj_m}(\tau,\cdot)\in\Ce^{N+1-k}(\R^d)$, 
justifying optimally the Taylor expansion used in order to obtain
\eqref{XNAalphagamma}.
{\bf (b)} 
Using 
the bounds implied by Assumption 
\ref{regularityassumption}, and the 
constants of Lemma \ref{DCNT}
and of Sobolev's imbedding theorem on $\R^d$,
the constant $C_r$ in \eqref{estres} could be given explicitly
via \eqref{esttN+1}, \eqref{estsN+1}, \eqref{estVWtilde}, \eqref{condalpha}.
%
}\end{remark}
%
\begin{remark}\label{rem_regularityassumption}{\em
%
%
 According to 
 \eqref{macro2}, \eqref{macro4}
 and definition 
 \eqref{source} of $\calS_{k-2,\whj_m}$,
%
using the property
$A,B\in\Ce\left([0,\tau_0];\Ha^s(\R^d)\right) \Rightarrow 
AB\in\Ce\left([0,\tau_0];\Ha^s(\R^d)\right)$ for $s>d/2$ 
(cf., e.g., \cite[Th.\ 4.39]{AdaFou02SoSp}),
it follows that 
 Assumption \ref{regularityassumption}
 is satisfied iff 
 \begin{align}\label{semilinearregularity}
 A_k
 \in\Ce^\lambda\left([0,\tau_0];\big(\Ha^{s+|\sigma|}(\R^d;\C)\big)^\nu\right)
 &\quad\text{for $\lambda+|\sigma|\le N+1-k$, $k=1,\ldots,N-1$,}
 \end{align}
 where $A_k:=(A_{k,1},\ldots,A_{k,\nu})^T$
(cf.\ for $N=3$ the examples in Sec.\ \ref{singleclosed3}, \ref{433} below).  
 Recall here, that 
 $A_N$ 
 remains undetermined by the formal derivation procedure, 
 and can thus be assumed as identically vanishing.

The determining equations \eqref{macro1} for $A_1$ 
and \eqref{macro3} for $A_k$, $k=2,\ldots,N-1$, 
are the (semilinear for $k=1$, linear inhomogeneous for $k=2,\ldots,N-1$) 
symmetric hyperbolic systems 
\begin{equation}\label{symhypsystem}
\pl_\tau A_k(\tau)+\calA_k A_k(\tau)=f_k(\tau, A_k(\tau))
\end{equation}
where $\calA_k:D(\calA_k)\to X_k$ is the infinitesimal generator 
of a $C_0$ semigroup on $X_k:=\big(\Ha^{M_k-1}(\R^d;\C)\big)^\nu$ 
with $X_k\supset D(\calA_k)\supset\big(\Ha^{M_k}(\R^d;\C)\big)^\nu=:Y_k$, 
$M_k\ge 1$(cf., e.g., \cite[\S 11.3.1]{RenRog93IPDE}, 
\cite[\S II.2.9]{Goldst85SLOA}).
Since $f_1$ does not depend explicitly on $\tau$ 
and is quadratic, 
and hence locally Lipschitz continuous in $A_1$, 
for $M_1>d/2$ it holds $f_1:Y_1\to Y_1$,  
and we obtain by standard results of semigroup theory
(cf., e.g., \cite[Th.\ 6.1.7]{Pazy83SLOA}, \cite[\S II.1.3]{Goldst85SLOA}) 
that there exists a $\tau_\mathrm{max}\le\infty$ 
such that \eqref{symhypsystem} has for $k=1$ and $A_1(0,\cdot)\in Y_1$ 
a unique classical solution 
$A_{1}\in\Ce^1([0,\tau_0];X_1)\cap\Ce([0,\tau_0];Y_1)$ 
for $\tau_0<\tau_\mathrm{max}$.
%
%
For $k=2,\ldots,N-1$ we have
\begin{equation*}
f_k(\tau, A_k(\tau))=F(\tau) A_k(\tau)+G(\tau)\ol{A_k}(\tau)+g_k(\tau)
\end{equation*}
with 
$F,G\in\Ce\big([0,\tau_0];\big(\Ha^{M_1}(\R^d;\C)\big)^{\nu\times\nu}\big)$ 
and 
$g_k\in\Ce\left([0,\tau_0];\big(\Ha^{M_{k-1}-2}(\R^d;\C)\big)^\nu\right)$.
In order to obtain unique classical solutions 
$A_{k}\in\Ce^1([0,\tau_0];X_k)\cap\Ce([0,\tau_0];Y_k)$ 
of the initial-value problems \eqref{symhypsystem} 
with $A_k(0,\cdot)\in Y_k$,
we need that $f_k:[0,\tau_0]\times Y_k\to Y_k$ is continuous in $(\tau,A_k)$
and uniformly Lipschitz in $A_k$. 
Hence, we need $M_{k-1}\ge M_k+2$, 
and thus $M_k\ge M_{N-1}+2(N{-}k{-}1)$ for $k=1,\ldots,N{-}1$.
%
%
Moreover, 
according to \eqref{semilinearregularity},
we need $M_k\ge s+N+1-k$, and in particular $M_{N-1}\ge s+2$.
Hence, for $M_k=s+2(N{-}k)$ all conditions on $M_k$ are satisfied,
which means that assuming initial data 
$A_k(0,\cdot)\in\Ha^{s+2(N-k)}(\R^d;\C)$,
for the equations \eqref{macro1} and \eqref{macro3}
we can guarantee \eqref{semilinearregularity}, and thus 
Assumption \ref{regularityassumption}.

In particular it is necessary and sufficient to assume
$A_1(0,\cdot)\in\Ha^{s+2(N-1)}(\R^d;\C)$.
Then, we can assume $A_{k}(0,\cdot)=0$ for $k=2,\ldots,N-1$.
This does not yield identically vanishing $k$-th order amplitudes $A_{k}$,
due to the source terms in \eqref{macro3}.
Of course, also some of the first-order amplitudes can be chosen as initially
vanishing.  For example, setting  in the case of 
the three-wave interaction \eqref{3wi} 
$A_{1,2}(0,\cdot)=0$, and assuming that $A_{1,1}(0,\cdot)$ and 
$A_{1,3}(0,\cdot)$ have disjoint
supports, the amplitudes $A_{1,1}$ and $A_{1,3}$ will be transported by
their different group velocities $\nabla_\vth\Om(\vth_1)$ 
and $\nabla_\vth\Om(\vth_3)$.  
The moment they interact, i.e., when their supports intersect, the amplitude 
$A_{1,2}$ arises. 
Note, however that if two of the three amplitudes are assumed to vanish
initially, then the third one is just transported, and does not give rise to
other amplitudes, since it does not interact with other pulses.
(Confer 
also the discussion on generation of pulses in 
Sec.\ \ref{secexamples} below, 
in particular Sec.\ \ref{threepulsesclosedN2}, Case \ref{case3wi},
and Sec.\ \ref{singleclosed2}.)  
\em}\end{remark}
%
%
\subsection{The justification result}
Having obtained the estimate \eqref{estres} of the residuum, 
we are now able 
to establish
the main result of our work, 
namely the justification of the macroscopic equations 
\eqref{macro1}, \eqref{macro2}, \eqref{macro3}, \eqref{macro4} 
obtained by formal derivation in Section \ref{formalderiv},
as giving the effective dynamics for an arbirtrary number of 
amplitude-modulated pulses in multidimensional lattices 
with scalar displacement.
More precisely we show:
\begin{theorem}\label{justiftheorem}
Let $\Gamma$ be the $d$-dimensional lattice \eqref{Gamma} ($d\in\N$)
and let $V_\alpha,W\in\Ce^{N+2}(\R)$ ($\alpha\in\Gamma$, 
$N\in\N$, $N>1{+}\frac{d}2$) 
be the interaction and on-site potentials \eqref{VW}
of the microscopic model \eqref{M}
satisfying 
$V_\alpha(x)=V_{-\alpha}(-x)$, \eqref{SC}, \eqref{condalpha}.
%
Let $\{\pm(\vth_j,\om_j): j=1,\ldots,\nu\}$ 
be a set of $\nu\in\N$ different pulses, 
closed under interactions up to order $N$ 
according to Definition \ref{defclosure},
and let the functions $A_{k,\whj_m}:[0,\tau_0]\to\C$ 
of the approximation $X_N^{A,\ve}$, \eqref{XNA}, 
solve the 
macroscopic 
equations 
\eqref{macro1}, \eqref{macro2}, \eqref{macro3}, \eqref{macro4}
as described in Theorem \ref{formalderivation},
and satisfy Assumption \ref{regularityassumption} for some $\tau_0>0$.

Then, for each $c>0$ there exist $\ve_0,C>0$ such 
that for all $\ve\in(0,\ve_0)$  
and any solution $x$ of \eqref{M} with 
\begin{align}
\Big\|\big(x(0),\dot x(0)\big)
-\big(X_{N-1}^{A,\ve}(0),\dot X_{N-1}^{A,\ve}(0)\big)\Big\|
_{\ell^2{\times}\ell^2}
&\le c\ve^\beta,
\qquad\text{$\beta\in\big(1,N{-}\textstyle\frac{d}2\big]$,}
\label{assindat}
\\
\intertext{it holds}
\Big\|\big(x(t),\dot x(t)\big)
-\big(X_{N-1}^{A,\ve}(t),\dot X_{N-1}^{A,\ve}(t)\big)\Big\|
_{\ell^2{\times}\ell^2}
&\le C\ve^\beta
\quad\text{for}\quad t\in[0,\tau_0/\ve].
\label{asst}
\end{align}
\end{theorem}
\begin{remark}\label{remindat}{\em
In view of Remark \ref{rem_regularityassumption},
the Assumption \ref{regularityassumption} above can be replaced
by the assumption 
$A_{k,j}(0,\cdot)\in\Ha^{s+2(N-k)}(\R^d;\C)$ 
for $k=1,\ldots,N-1$, $j=1,\ldots,\nu$ with $s>d/2$ 
(and in particular by $A_{1,j}(0,\cdot)\in\Ha^{s+2(N-1)}(\R^d;\C)$, 
$A_{k,j}(0,\cdot)=0$, $k\ge 2$),
which moreover guarantees 
the solvability of the respective amplitude equations 
\eqref{macro1} for $k=1$ and \eqref{macro3} for $k\ge 2$
up to some $\tau_0>0$.
\em}\end{remark}
%
%
{\bf Proof:}
%
%
We write the microscopic model \eqref{M}
as a first order system in  
$Y:=\ell^2(\Gamma){\times}\ell^2(\Gamma)$,
\begin{equation}\label{sysx}
\dot{\wt x}=\wt \calL\wt x+\wt \calM(\wt x)
\quad\text{with}\quad
\wt x:=\begin{pmatrix} x \\ \dot x \end{pmatrix},\quad
\wt \calL:=\begin{pmatrix}0 & \calI \\ \calL & 0\end{pmatrix},\quad
\wt \calM(\wt x):=\begin{pmatrix} 0 \\ \calM(x) \end{pmatrix},
\end{equation}
where $\calI:\ell^2(\Gamma)\to\ell^2(\Gamma)$ is the identity, 
and $\calL,\calM:\ell^2(\Gamma)\to\ell^2(\Gamma)$ are given by
\begin{equation}
(\calL x)_\g:=
\sum_{\alpha\in\Gamma} a_{1,\alpha}(x_{\g+\alpha}{-}x_\g)-b_1x_\g,
\quad
(\calM(x))_\g:=
\sum_{\alpha\in\Gamma}V^\prime_{2,\alpha}(x_{\g+\alpha}{-}x_\g)
-W^\prime_{2}(x_\g)
\label{N}
\end{equation}
with 
$a_{1,\alpha}$, $V^\prime_{2,\alpha}$, 
$b_1$, $W^\prime_{2}$ as in 
\eqref{VW}.
%
%
On 
the Banach space 
$Y$ we use the 
energy
norm
\begin{equation*}
\|(x,y)\|_Y^2:=\|x\|_E^2+\|y\|_{\ell^2}^2
,\qquad
\|x\|_E^2:=
\sum_{\alpha\in\Gamma}\frac{a_{1,\alpha}}2
\sum_{\g\in\Gamma}|x_{\g+\alpha}{-}x_\g|^2
+b_1\sum_{\g\in\Gamma}|x_\g|^2,
\end{equation*}
defined in such a way that its square is twice the 
harmonic part of the Hamiltonian $\calH$, cf.\ \eqref{KU}.
%
As is well known, 
the flow of the linearized system $\dot{\wt x}=\wt\calL\wt x$
preserves this norm,
i.e.\ its associated semigroup $\e^{t\wt\calL}$   
satisfies $\|\e^{t\wt \calL}\|_{Y\to Y}=1$
(cf., e.g., \cite[Prop.\ 3.1]{GM04}). 
%
 Moreover, since $\|\cdot\|_{\ell^2}$ and $\|\cdot\|_E$ are equivalent 
 by the stability assumption \eqref{SC}, 
 \begin{equation*}
\mu_-\|x\|_{\ell^2}\le\|x\|_E\le \mu_+\|x\|_{\ell^2},
\qquad 0<\mu_-:=\min\limits_{\vth\in\tor}\Om(\vth),
\quad \mu_+:=\max\limits_{\vth\in\tor}\Om(\vth),
 \end{equation*}
which follows by Fourier transformation,
%
%
we obtain the equivalence of 
$\|\cdot\|_{\ell^2\times\ell^2}$ and $\|\cdot\|_Y$, 
 \begin{equation}\label{normequivY}
 \wt\mu_-\|\wt x\|_{\ell^2\times\ell^2}\le\|\wt x\|_Y
 \le \wt\mu_+\|\wt x\|_{\ell^2\times\ell^2},
 \qquad \wt\mu_-:=
 \min\{\mu_-,1\},
 \quad \wt\mu_+:=
 \max\{\mu_+,1\}.
 \end{equation}
%
%

We show that the {\it error} 
$\wt R_\ve(t):=\ve^{-\beta}\big(\wt x(t)-\wt X_{N}^{A,\ve}(t)\big)$
between a solution $\wt x$ of 
(\ref{sysx}) 
and the approximation 
$\wt X_{N}^{A,\ve}:=\Big(X_{N}^{A,\ve},\dot X_{N}^{A,\ve}\Big)^T$ 
satisfies
\begin{equation}\label{errorestimate}
\big\|\wt R_\ve(0)\big\|_Y\le c_0
\quad\Longrightarrow\quad
\big\|\wt R_\ve(t)\big\|_Y\le D
\quad\text{for $\ve\le\ve_0$, $\ve t\le \tau_0$.} 
\end{equation}
By 
\eqref{normequivY},
this is the assertion of the theorem
with $\wt X_{N}^{A,\ve}$ instead of $\wt X_{N-1}^{A,\ve}$.
%
However, according to 
\eqref{XNA}, 
Lemma \ref{DCNT}
and Assumption \ref{regularityassumption},
there exists a $c_N>0$ such that
\begin{align*}
\big\|\wt X_{N}^{A,\ve}(t){-}\wt X_{N-1}^{A,\ve}(t)\big\|_Y
&
 \le\ve^N\sum_{\whj_m\in\at_N}
\!
 \Big(\mu_{+}^2\big\|A_{N,\whj_m}\big\|_{\ell^2}^2
 +\big(\ve\big\|\partial_\tau A_{N,\whj_m}\big\|_{\ell^2}
 +|\om_{\whj_m}|\big\|A_{N,\whj_m}\big\|_{\ell^2}\big)^2\Big)^{\frac12}
%
%
\notag
\\&
\le\ve^{N-d/2}c_N
\quad\text{for $\ve\le\ve_0$, $\ve t\le \tau_0$.}
\end{align*} 
%
Hence, by \eqref{normequivY} and $\beta\le N-d/2$,
we obtain from \eqref{assindat}
\begin{align*}
\big\|\wt R_\ve(0)\big\|_Y
&\le 
\ve^{-\beta}\big\|\wt x(0){-}\wt X_{N-1}^{A,\ve}(0)\big\|_Y
+\ve^{-\beta}\big\|\wt X_{N}^{A,\ve}(0){-}\wt X_{N-1}^{A,\ve}(0)\big\|_Y
\notag\\&
\le \wt\mu_+c+\ve_0^{N-d/2-\beta}c_N=:c_0 
\quad\text{for $\ve\le\ve_0$.}
\end{align*}
%
%
Vice versa, 
from the right hand side of \eqref{errorestimate}
and $\beta\le N-d/2$
we obtain
\eqref{asst}
\begin{align*}
\ve^{-\beta}\big\|\wt x(t){-}\wt X_{N-1}^{A,\ve}(t)\big\|_{\ell^2\times\ell^2}
&
\le  \wt\mu_-^{-1}\left(
\big\|\wt R_\ve(t)\big\|_Y
+\ve^{-\beta}\big\|\wt X_{N}^{A,\ve}(t){-}\wt X_{N-1}^{A,\ve}(t)\big\|_Y
\right)
\\&
\le  \wt\mu_-^{-1}(D+\ve_0^{N-d/2-\beta}c_N)=:C
\quad\text{for $\ve\le\ve_0$, $\ve t\le \tau_0$.}
\end{align*}
%

It remains to prove \eqref{errorestimate}.
Inserting $\wt R_\ve$ into (\ref{sysx})
we obtain the differential equation
\begin{align*}
\dot{\wt R}_\ve
&=
\wt \calL\wt R_\ve
+\ve^{-\beta}\left(\wt \calM\big(\wt X_{N}^{A,\ve}+\ve^\beta\wt R_\ve\big)
-\wt \calM\big(\wt  X_{N}^{A,\ve}\big)-\res\big(\wt X_{N}^{A,\ve}\big)\right)
\end{align*}
with $\ds \res\big(\wt X_{N}^{A,\ve}\big)
:=\dot{\wt X}{}_{N}^{A,\ve}
-\wt \calL\wt X_{N}^{A,\ve}-\wt \calM\big(\wt  X_{N}^{A,\ve}\big)$,
and write it in its integral form
\begin{equation*}
\wt R_\ve(t)=\e^{t\wt \calL}\wt R_\ve(0)
+\ve^{-\beta}\int_0^t \e^{(t{-}s)\wt \calL}
\Big(\wt \calM\big(\wt X_{N}^{A,\ve}(s){+}\ve^\beta\wt R_\ve(s)\big)
-\wt \calM\big(\wt X_{N}^{A,\ve}(s)\big)
-\res\big(\wt X_{N}^{A,\ve}(s)\big)\Big)\de s,
\end{equation*}
%
%
$\e^{t\wt \calL}$ the semigroup 
to 
$\dot{\wt R}_\ve=\wt \calL\wt R_\ve$.
Since $\|\e^{t\wt\calL}\|_{Y\to Y}=1$ (cf.\ above),
taking the norm gives
\begin{multline}\label{varconst}
\big\|\wt R_\ve(t)\big\|_Y
\le\big\|\wt R_\ve(0)\big\|_Y
+\ve^{-\beta}\int_0^t\Big( 
\big\|\wt \calM\big(\wt X_{N}^{A,\ve}(s){+}\ve^\beta\wt R_\ve(s)\big)
-\wt \calM\big(\wt X_{N}^{A,\ve}(s)\big)\big\|_Y
\\
+\big\|\res\big(\wt X_{N}^{A,\ve}(s)\big)\big\|_Y\Big)\de s.
\end{multline}
%
%

We estimate the norms on the right hand side of \eqref{varconst}.
According to 
the formal derivation in 
Section \ref{formalderiv},
%
$\ddot{X}_N^{A,\ve}-\calL X_N^{A,\ve}-\calM\big(X_N^{A,\ve}\big)
=\res\big(X_N^{A,\ve}\big)$.
Hence, 
Lemma \ref{lemmares} 
gives 
\begin{align}\label{estresY}
\big\|\res\big(\wt X_N^{A,\ve}\big)\big\|_Y
=\big\|\res\big(X_N^{A,\ve}\big)\big\|_{\ell^2}
\le \ve^{N+1-d/2} C_r
\quad\text{for $\ve\in(0,\ve_0]$, $t\in[0,\tau_0/\ve]$.}
\end{align}
Next, we estimate the norm of the nonlinear terms. 
%
%
From \eqref{N} we obtain 
\begin{equation*}
\left\|\calM(x)-\calM(y)\right\|_{\ell^2}
\le
\sum_{\alpha\in\Gamma}
\left\|V^\prime_{2,\alpha}(x_{\cdot+\alpha}{-}x)
-V^\prime_{2,\alpha}(y_{\cdot+\alpha}{-}y)\right\|_{\ell^2}
+\left\|W^\prime_{2}(x)-W^\prime_{2}(y)\right\|_{\ell^2}
\end{equation*}
for $x,y\in\ell^2(\Gamma)$, 
where we denote $(x_{\cdot+\alpha})_\g:=x_{\g+\alpha}$ 
and $(V(x))_\g:=V(x_\g)$ for $V:\R\to\R$.
%
%
Since $V_{\alpha}, W\in\Ce^{N+2}(\R)$ with $N\ge2$
and 
$V^{\prime\prime}_{2,\alpha}(x)=\OO(|x|)$,
$W^{\prime\prime}_2(x)=\OO(|x|)$
(cf.\ \eqref{VW}),
we obtain by the mean value theorem
\begin{align*}
&
\big|V^\prime_{2,\alpha}(x)-V^\prime_{2,\alpha}(y)\big|
\le
c_{\alpha,x_0}
\big(|x|{+}|y|\big)|x-y|,
\quad
\big|W^\prime_2(x)-W^\prime_2(y)\big|
\le
c_{x_0}
\big(|x|{+}|y|\big)|x-y|
\end{align*}
for $x,y\in[-2x_0,2x_0]$, $x_0>0$, with 
$c_{\alpha,x_0},c_{x_0}>0$.
%
Hence, there exists a $c_{\calM,x_0}>0$ (cf.\ also \eqref{condalpha}), 
such that for $\wt x,\wt y\in Y$
\begin{equation}\label{estNgeneral}
\big\|\wt \calM(\wt x)-\wt \calM(\wt y)\big\|_Y
\le c_{\calM,x_0}
\left(\|x\|_{\ell^\infty}{+}\|y\|_{\ell^\infty}\right)\|\wt x-\wt y\|_Y
\quad\text{for $\|x\|_{\ell^\infty},\,\|y\|_{\ell^\infty}\le x_0$.}
\end{equation}
Now, we apply \eqref{estNgeneral} on 
$\wt x:=\wt X_{N}^{A,\ve}+\ve^\beta\wt R_\ve$ and $\wt y:=\wt X_{N}^{A,\ve}$.
By \eqref{estX} 
%
%
and Assumption \ref{regularityassumption} 
there exists a $c_A>0$ such that
\begin{align*}
\big\|X_N^{A,\ve}\big\|_{\ell^\infty}
\le\ve 
c_A
\quad\text{for $\ve\le\ve_0$, $\ve t\le \tau_0$.} 
\end{align*}
%
We set 
$D:=\big(c_0+\ve_0^{N-d/2-\beta}\,\tau_0\, C_r\big)\e^{\tau_0 C_\calM}$
with $c_0,C_r$ as in \eqref{errorestimate}, \eqref{estresY},
$C_\calM:=3c_{\calM,x_0}c_A$,
and $\ve_0>0$ such that 
$\ve_0 c_A\le x_0/2$
%
%
and $\ve_0^{\beta-1}D/\mu_{-}\le c_A$, 
thereby assuming $\beta>1$.
%
%
%
Since 
$\big\|\wt R_\ve(0)\big\|_Y\le c_0<D$
and $\big\|\wt R_\ve(t)\big\|_Y$ is continuous,
there exists for every $\ve\in(0,\ve_0]$ a $t_D^\ve>0$, such that
$\big\|\wt R_\ve(t)\big\|_Y\le D$ for $t\le t_D^\ve$.
Then,
it holds
\begin{align*}
\big\|X_N^{A,\ve}\big\|_{\ell^\infty}
,\ 
\big\|X_N^{A,\ve}+\ve^\beta R_\ve\big\|_{\ell^\infty}
\le x_0
\quad\text{for $\ve\le\ve_0$, 
$t\le \min\{\tau_0/\ve,t_D^\ve\}$},
\end{align*}
and \eqref{estNgeneral} gives
\begin{equation*}
\big\|\wt \calM\big(\wt X_{N}^{A,\ve}{+}\ve^\beta\wt R_\ve\big)
-\wt \calM\big(\wt X_{N}^{A,\ve}\big)\big\|_Y
\le \ve^{\beta+1}C_\calM\big\|\wt R_\ve\big\|_Y
\quad\text{for $\ve\le\ve_0$, 
$t\le \min\{\tau_0/\ve,t_D^\ve\}$.}
\end{equation*}
Inserting this estimate, \eqref{estresY}
and $\big\|\wt R_\ve(0)\big\|_Y\le c_0$ into \eqref{varconst}, we obtain
\begin{equation*}
\big\|\wt R_\ve(t)\big\|_Y
\le 
\Big(c_0+\ve_0^{N-d/2-\beta}\,\tau_0\, C_r\Big)
+\ve\,C_\calM \int_0^t\big\|\wt R_\ve(s)\big\|_Y \de s
\quad\text{for $\ve\le\ve_0$, 
$t\le \min\{\tau_0/\ve,t_D^\ve\}$},
\end{equation*}
and Gronwall's lemma gives
\begin{align*}
\big\|\wt R_\ve(t)\big\|_Y\le
\Big(c_0+\ve_0^{N-d/2-\beta}\, \tau_0\, C_r\Big)
\e^{\ve t C_\calM}
\quad\text{for $\ve\le\ve_0$, 
$t\le \min\{\tau_0/\ve,t_D^\ve\}$.}
\end{align*}
Hence, $\big\|\wt R_\ve(t)\big\|_Y\le D$ 
for $\ve\le\ve_0$, $t\le\tau_0/\ve\le t_D^\ve$,
and the proof of \eqref{errorestimate}, 
and thus of the theorem, is completed.\hfill$\square$ 
%
%
%
\section{Examples}\label{secexamples}
In the present section we give, on the one hand, a complete classification of 
all possible types of macroscopic systems for the first-order
amplitudes $A_{1,j}$ of $\nu=1,2,3$ modulated {\it different} pulses, 
which form sets $\{\pm(\vth_j,\om_j):\ j=1,\ldots,\nu\}$ 
closed under interactions up to order $N=2$.
The classification for a given number of pulses results from the number and
type of resonances 
(of order $2$)
between them.
These resonances can be traced back on the coupling terms of the 
first-order amplitude equations.
Each case leads to a different
macroscopic system, which can not be obtained from another even when the
resonance conditions for the former are a special case of those for the
latter (cf.\ the discussion in Sec.\ \ref{threepulsesclosedN2}, 
Case \ref{ca4}).
In particular, additional resonances 
lead to additional terms in the
macroscopic systems 
(cf., e.g., in Sec.\ \ref{threepulsesclosedN2} the Cases 
\ref{ca3} with \ref{ca4} or \ref{case3wi}, \ref{ca5},
and \ref{case3wi} with that of Sec.\ \ref{432}). 
Naturally, the more pulses we consider the more different constellations
of resonances can appear, which of course can include also closed subsystems
(cf.\, e.g., \eqref{singletransport} with \eqref{twononinterpulses}
or Sec.\ \ref{threepulsesclosedN2}, Case \ref{ca1},
and \eqref{nu2selfamp} with Sec.\ \ref{threepulsesclosedN2}, Case \ref{ca3}).
More precisely, 
{\it within} a system of $\nu$ pulses 
with 
$0<\om_1<\ldots<\om_\nu$
the resonances
of order $2$ 
\begin{equation*}
(\vth_j+\vth_i,\om_j+\om_i)=(\vth_k,\om_k),
\quad k>j,i,
\quad k,j,i=1,\ldots,\nu,
\end{equation*}
can appear. 

On the other hand, we exemplify on the considered sets the essence of the
closedness condition. Only when it is satisfied we are able to obtain 
complete effective dynamics, in the sense that only then we can choose 
(except for regularity restrictions)
arbitrary initial data for the macroscopic amplitudes of the pulses
(cf.\  in particular the discussions of non-$2$-closed sets in 
Sec.\  \ref{singlenonclosed2} and \ref{twopulsesnonclosedN2}, and of
non-$3$-closed ones in Sec.\ \ref{singleclosed3}).
%

Recall that whenever in the following 
a set is closed up to interactions of order $N=2$ 
($2$-closed) 
the derived macroscopic equations 
provide us with the first-order amplitudes $A_{1,j}$ and the functions 
$A_{2,\whj_m}$, 
$\whj_m\in\at_2\setminus\calN$ 
of the second-order approximation $X_2^{A,\ve}$
such that $\res\big(X_2^{A,\ve}\big)=\OO(\ve^3)$,
while the second-order amplitudes $A_{2,j}$ remain undetermined, 
enabling us to choose $A_{2,j}=0$.
%
Thus, according to Theorem \ref{justiftheorem} and Remark \ref{remindat}, 
for one-dimensional lattices (oscillator chains) with $V_\alpha,W\in\Ce^4(\R)$
the given macroscopic equations for $A_{1,j}$ 
with $A_{1,j}(0,\cdot)\in\Ha^{3}(\R)$ 
establish the effective dynamics of the corresponding modulated pulses
in the sense of Theorem \ref{justiftheorem}, 
i.e.\ with $N-1=1$ and $\beta\in(1,3/2]$ in \eqref{assindat} and \eqref{asst}
and for macroscopic time intervalls $[0,\tau_0]$, 
where $\tau_0>0$ is limited only by the time of existence of the macroscopic
solutions $A_{1,j}$.

For deriving valid effective dynamics 
in multidimensional lattices with $d\ge 2$, however,
we need the considered set of pulses to be closed 
under interactions up to order $N>1+d/2$. 
In particular for $d=2$ or $3$ we need $N=3$.
Then, for $V_\alpha,W\in\Ce^5(\R)$, 
the equations determining the macroscopic coefficients of $X_2^{A,\ve}$
with 
$A_{1,j}(0,\cdot)\in\Ha^6(\R^d)$ 
and 
$A_{2,j}(0,\cdot)\in\Ha^4(\R^d)$ 
(or $A_{2,j}(0,\cdot)=0$)
give the effective dynamics of the corresponding modulated pulses 
in the sense of Theorem \ref{justiftheorem}, 
i.e.\ with $N-1=2$ and $\beta\in(1,2]$ or $\beta\in(1,3/2]$, respectively,
in \eqref{assindat} and \eqref{asst}. 
We give in the following the corresponding macroscopic equations 
in the case of a single pulse
(Sec.\ \ref{singleclosed2}, Sec.\ \ref{singleclosed3})
and in the case of a three-wave-interaction 
(Sec.\ \ref{threepulsesclosedN2}, Case \ref{case3wi}, Sec.\ \ref{433}).

We would like to remark that of course 
each single model 
presented in the following
could be discussed and interpreted 
more extensively than done here.
%
Note, in this context 
that the presented macroscopic equations are 
(at first formally)
valid for pair-interaction potentials among atoms at an arbitrary distance
in lattices of arbitrary dimension.  
Specifying the potentials 
(as for instance by considering only nearest-neighbour interactions) 
and the lattice dimension would lead naturally to more 
concrete results and possible interpretations.
Nevertheless, we believe that the comparative presentation of all these models
is helpfull in getting an impresion of the variety of possible 
interactions of pulses, 
and even may serve as a reference for further investigations.

Concerning the following examples the most crucial question is of course 
whether the resonance 
conditions corresponding to each model can be satisfied at all 
for a given potential in a given dimension.   
Here, we restrict ourselves in addresing this question 
exemplarily 
in Sec.\ \ref{interactionexistence}
for the case of a three-wave-interaction
in an oscillator chain ($d=1$) with nearest-neighbour interaction potential
and a stabilizing on-site potential.
This example shows that the satisfaction of resonance conditions 
is equivalent to the solution of algebraic equations in $n$ variables,
where $n$ depends on the number of independent pulses, cf.\ also 
\cite[Sec.\ III.E]{GHM08a}.
In the present case $n=2$, and it turns out that a three-wave-interaction can
appear only when the leading, harmonic part of the interaction
potential is repulsive.  
Recall here, that the satisfaction of resonance and nonresonance conditions
depends only on the harmonic parts of the involved potentials, since they are
linked to the dispersion function of the (linearized) lattice, see \eqref{DR}
and Definitions \ref{resonance} and \ref{defclosure}.
Moreover, note that, since closedness (nonresonance) 
conditions appear as avoided solutions of algebraic equations, as
shown in our example, they are usually more easily satisfied than resonance
conditions. 
%
%
However, aside from the observation that the latter depend on the number of
involved pulses, the range of the interaction potential and the dimension of
the lattice, we do not follow here these questions further. 
%
%
\subsection{A single pulse}
We start with the case $\nu=1$ of a single pulse $\pm(\vth_1,\om_1)$ 
with $\om_1=\Om(\vth_1)>0$
and consider the second-order approximation \eqref{XNA}
\begin{equation}\label{nu1N2}
X_2^{A,\ve}
=\ve A_{1,1}\boe_1+\ve^2\Big(A_{2,1}\boe_1
+A_{2,(1,1)}\boe_1^2
+{\frac12}A_{2,(1,-1)}
\Big)+\cc
\end{equation}
Here, $\boe_{\pm1}(t,\g)=\e^{\i(\om_{\pm1} t+\vth_{\pm1}\cdot\g)}$ 
with $\om_{\pm1}=\pm\om_1$, $\vth_{\pm1}=\pm\vth_1$.
The multi-index $(1,-1)\in\calN^2$ is the representant of $\pm(1,-1)$. 
%
The factor $\frac12$ in front of $A_{2,(1,-1)}\in\R$ 
arises, since we abbreviate the complex conjugates of all explicitly written 
functions on the right hand side of \eqref{nu1N2} by $\cc$
%
\subsubsection{2-closed system of a single pulse}
\label{singleclosed2}
According to Definition \ref{defclosure}, 
the set of pulses  $\{\pm(\vth_1,\om_1)\}$ is closed under interaction
up to order $2$ if
\begin{center}
\emph{the pair $(2\vth_1,2\om_1)$ 
does not characterize a pulse. 
} 
\end{center}
Recall here, that the pair $(0,0)$ corresponding to the functions
$\boe_{\pm(1,-1)}=\boe_{\pm1}\boe_{\mp1}=1$ does not characterize a pulse 
by our stability assumption \eqref{SC}: $\Om^2(\vth)>0$ 
for all $\vth\in\tor$.
In particular, $(\vth_1,\om_1)\neq (0,0)$,
which yields $(2\vth_1,2\om_1)\neq\pm(\vth_1,\om_1)$,
meaning that the functions $\boe_{\pm1}^2$ generated by 
the pulses $\boe_{\pm1}$ can not equal one of the latter. 
Hence, here the closedness condition is equivalent to 
the nonresonance condition
\begin{equation*}
\delta_{(1,1)}=\Om^2(2\vth_1)-4\om_1^2\neq 0,
\end{equation*}
and we can determine
the function $A_{1,1}$ in \eqref{nu1N2}
by the equation \eqref{macro1}
\begin{equation}\label{singletransport}
\partial_\tau A_{1,1}
-\nabla_\vth\Omega(\vth_1){\cdot}\nabla_y A_{1,1}=0,
\end{equation}
and the functions $A_{2,(1,\pm1)}$ by the equations \eqref{macro2}
 \begin{align*}
 A_{2,(1,1)}=\frac{c_{(1,1)}}{\delta_{(1,1)}}A_{1,1}^2,
 \qquad
 A_{2,(1,-1)}=
 \frac{-2b_2}{b_1}
 |A_{1,1}|^2
 \end{align*}
with 
\begin{align}\label{c11}
c_{(1,1)}&=
-4\i\sum_{\alpha\in\Gamma}a_{2,\alpha}
\sin^2\left(\frac{\vth_1}2{\cdot}\alpha\right)
\sin\left(\vth_1{\cdot}\alpha\right)
-b_2.
\end{align}

According to \eqref{singletransport},
the amplitude $A_{1,1}$ of a single modulated pulse, 
which does not generate a further pulse via self-interaction,
is simply transported with its group velocity $\nabla_\vth\Omega(\vth_1)$. 
%
Note here, that this does not mean that the pulse $\pm(\vth_1,\om_1)$ 
could not possibly interact with some other pulse $\pm(\vth_2,\om_2)$. 
It just means that if initially all other pulses have vanishing amplitude,
they cannot be generated (i.e., exhihit at a later time a non-vanishing 
amplitude) by a single pulse, except in the case of self-interaction, 
$\delta_{(1,1)}=0$, 
which is discussed in the following. 
%
\subsubsection{Non-2-closed system of a single pulse}
\label{singlenonclosed2}
The closedness condition for the system of a single pulse is violated 
if $\delta_{(1,1)}=0$, i.e., in the case where the pulse $\pm(\vth_1,\om_1)$
generates via self-interaction a further pulse $\pm(2\vth_1,2\om_1)$.
Considering still, by the ansatz \eqref{nu1N2}, 
\emph{only} the pulse $\pm(\vth_1,\om_1)$ we would obtain $A_{1,1}=0$ 
(cf.\ the equation for $A_{2,(1,1)}$ in Sec.\ \ref{singleclosed2} above).
This means that we would not consider any modulated pulses at all, 
and in particular 
that non-vanishing initial data $A_{1,1}(0,\cdot)\neq0$ would be excluded.
The situation is resolved if we take into account \emph{also} 
(at least) the generated pulse $(\vth_2,\om_2):=(2\vth_1,2\om_1)$.
Since $(\vth_2,\om_2)\neq \pm(\vth_1,\om_1)$, 
this leads to the case of a set of $\nu\ge2$ pulses, 
which for $\nu=2,3$ is discussed in 
Sections \ref{twopulses} and \ref{threepulses} respectively.

\subsubsection{3-closed system of a single pulse}
\label{singleclosed3}
The third-order approximation \eqref{XNA} for a single pulse 
$\pm(\vth_1,\om_1)$ with $\om_1=\Om(\vth_1)>0$
is given by 
\begin{equation}\label{nu1N3}
X_3^{A,\ve}=X_2^{A,\ve}
+\ve^3\Big(A_{3,1}\boe_1+A_{3,(1,1)}\boe_1^2+A_{3,(1,1,1)}\boe_1^3
+{\frac12}A_{3,(1,-1)}+\cc\Big)
\end{equation}
where $X_2^{A,\ve}$ is the second-order approximation \eqref{nu1N2}.
Here, the set of representants is
$\at_3=\{\pm 1, \pm (1,1), \pm(1,1,1), (1,-1)\}$ with  
$\at_2=\{\pm (1,1), (1,-1)\}$ and $\calN=\{\pm 1\}$.
Hence, $\calN$ is 
closed under interactions up to order $3$
if the nonresonance conditions
\begin{equation*}
\delta_{(1,1)}=\Om^2(2\vth_1)-4\om_1^2\neq 0,
\quad  
\delta_{(1,1,1)}=\Om^2(3\vth_1)-9\om_1^2\neq 0 
\end{equation*}
are satisfied.
The second-order amplitude $A_{2,1}$ is determined by 
equation 
\eqref{macro3} for $k=3$
\begin{multline*}
\partial_\tau A_{2,1}-\nabla_\vth\Omega(\vth_1){\cdot}\nabla_y A_{2,1}
=\frac1{2\i\Omega(\vth_1)}
\Bigg(\Big(2\frac{|c_{(1,1)}|^2}{\delta_{(1,1)}}
+4\frac{b_2^2}{b_1}+3c_{(1,1,-1)}\Big)|A_{1,1}|^2A_{1,1}
\\
+\frac12\sum_{\alpha\in\Gamma}a_{1,\alpha}
\e^{\i\vth_1\cdot\alpha}(\alpha{\cdot}\nabla_y)^2 A_{1,1}
-\partial_\tau^2 A_{1,1}\Bigg),
\end{multline*}
and equation \eqref{macro4} for $k=3$ gives
\begin{align*}
b_1 A_{3,(1,-1)}
&=
-2b_2 A_{1,1}
\ol{A_{2,1}}
+2A_{1,1}\sum_{\alpha\in\Gamma}
a_{2,\alpha}\left(1{-}\e^{-\i\vth_1\cdot\alpha}\right)\alpha{\cdot}\nabla_y 
\ol{A_{1,1}}
+\cc,
\\
\delta_{(1,1)} A_{3,(1,1)}
&=
\frac{c_{(1,1)}}{\delta_{(1,1)}}\big(-4\i\om_1\partial_\tau 
+2\i\Omega(2\vth_1)\nabla_\vth\Omega(2\vth_1){\cdot}\nabla_y\big)A_{1,1}^2
+2c_{(1,1)}A_{1,1}A_{2,1}
\\&\quad
+2\sum_{\alpha\in\Gamma}a_{2,\alpha}
\left(\e^{2\i\vth_1\cdot\alpha}{-}\e^{\i\vth_1\cdot\alpha}\right)
\alpha{\cdot}A_{1,1}\nabla_y A_{1,1},
\\
\delta_{(1,1,1)} A_{3,(1,1,1)}
&=
\Big(2c_{(1,(1,1))}\frac{c_{(1,1)}}{\delta_{(1,1)}}
+c_{(1,1,1)}\Big)A_{1,1}^3,
\end{align*}
where we used the functions $A_{2,(1,1)}$ and $A_{2,(1,-1)}$
calculated in Sec.\ \ref{singleclosed2}, 
and where $c_{(1,1)}$, $c_{(1,(1,1))}$, and $c_{(1,1,1)}$, $c_{(1,1,-1)}$
are given by \eqref{c11}, \eqref{c2gen}, and \eqref{c}, respectively.
Hence, the second-order amplitude $A_{2,1}$ is determined by an 
inhomogeneous transport equation, where the source term is determined by 
the first-order amplitude $A_{1,1}$ calculated by \eqref{singletransport}.
Then, since $\delta_{(1,-1)},\delta_{(1,1)},\delta_{(1,1,1)}\neq 0$ by the
nonresonance conditions, and knowing the amplitudes $A_{1,1}$, $A_{2,1}$,
we can calculate the functions $A_{3,(1,\pm1)}$, $A_{3,(1,1,1)}$.
The third-order amplitude $A_{3,1}$ remains undetermined. 
Setting $A_{3,1}=0$, 
and inserting the above 
functions and $X_2^{A,\ve}$ 
from Sec.\ \ref{singleclosed2} into \eqref{nu1N3}, 
we obtain 
$X_3^{A,\ve}$ with $\res\big(X_3^{A,\ve}\big)=\OO(\ve^4)$.

Note here, that this derivation of amplitude equations for $A_{2,1}$ 
and $A_{3,(1,\pm1)}$ is possible only since we anticipated for functions 
$A_{k,\whj_m}$ with $\whj_m\in\at_k$ and not only with $\whj_m\in\calN^k$
for $k=2,3$ in \eqref{XNA}. 
In the latter case, we would have assumed 
$A_{2,1}=A_{3,(1,\pm1)}=0$ identically, 
and would obtain instead of the above equations 
for $A_{2,1}$ and $A_{3,(1,\pm1)}$ additional conditions on the fist-order
amplitude $A_{1,1}$ which would at least restrict our choice of initial
conditions. 

The $3$-closedness of the system 
would be violated if $\delta_{(1,1,1)}=0$, i.e., if $(3\vth_1,3\om_1)$ is a
pulse (resonance of order $3$), not considered by our system of a single pulse.
Ignoring this generated pulse would give $A_{1,1}=0$ identically (cf.\ the
equation for $A_{3,(1,1,1)}$), prohibiting non-trivial dynamics even for the
{\it first-order} amplitude $A_{1,1}$.
Again, the remedy consists in taking into account also the generated pulse,
thus considering a set of two pulses.  
Note, however, that the resonance $\delta_{(1,1,1)}=0$ is detected only by the
third-order approximation $X_3^{A,\ve}$.
Thus, even in its presence the system remains $2$-closed 
(when $\delta_{(1,1)}\neq 0$), 
and the derivation of valid effective dynamics of the first-order amplitude 
$A_{1,1}$ is still possible for one-dimensional lattices. 
This is in line with the order $\beta\in(1,3/2]$ of the error in  
\eqref{assindat} and \eqref{asst}, 
which for $N=2$ is too coarse to detect this interaction.
Of course, if one is interested in valid higher-order effective dynamics
(with $N\ge 3$) the same problems as discussed above arise.
%
\subsection{Two pulses}
\label{twopulses}
We consider the case $\nu=2$ of two different pulses 
$(\vth_1,\om_1)\neq(\vth_2,\om_2)$ with $\om_j=\Om(\vth_j)>0$, $j=1,2$.
The second-order approximation \eqref{XNA} 
reads
\begin{multline}
X_2^{A,\ve}
=\ve\sum_{j=1}^2A_{1,j}\boe_j
+\ve^2\Big(\sum_{j=1}^2 \big(A_{2,j}\boe_j+A_{2,(j,j)}\boe_j^2\big)
+A_{2,(1,2)}\boe_1\boe_2+A_{2,(1,-2)}\boe_1\boe_{-2}
\\
+{\frac12} A_{2,(1,-1)}
\Big)+\cc,
\label{nu2N2}
\end{multline}
with $\boe_j(t,\g)=\e^{\i(\om_j t+\vth_j\cdot\g)}$, $j=\pm1,\pm2$, 
where $\om_{-j}=-\om_{j}$,  $\vth_{-j}=-\vth_{j}$.
Here, $(1,2)$, $(1,-2)$ represent also $(2,1)$, $(-2,1)$, respectively, 
and $(1,-1)$ represents $\pm(1,-1),\pm(2,-2)$. 
%
\subsubsection{2-closed systems of two pulses}
\label{twopulsesclosedN2}
According to Definition \ref{defclosure},
the set of pulses  $\{\pm(\vth_j,\om_j): j=1,2\}$ is closed under inter\-action
up to order $2$ if 
\begin{center}
\emph{the pairs $(2\vth_j,2\om_j)$, $j=1,2$, 
$(\vth_1{\pm}\vth_2,\om_1{\pm}\om_2)$
do not characterize pulses, 
\\
except if either 
$(2\vth_1,2\om_1)=(\vth_2,\om_2)$ or $(2\vth_2,2\om_2)=(\vth_1,\om_1)$.} 
\end{center}
The presence or not of one of these two exceptional cases, leads to two
different cases among $2$-closed sets of two pulses:\\

\noindent{\bf Case 1: Non-interacting pulses.}\\[2mm]
Excluding the exceptional cases above, 
the closedness condition is equivalent to the nonresonance conditions
\begin{equation}\label{nu2N2clcond}
\delta_{(j,j)}=\Om^2(2\vth_j)-4\om_j^2\neq 0,\quad j=1,2,\qquad
\delta_{(1,\pm2)}=\Om^2(\vth_1{\pm}\vth_2)-(\om_1{\pm}\om_2)^2\neq 0,
\end{equation}
and we can determine the first-order amplitudes $A_{1,j}$
in \eqref{nu2N2} by the equations \eqref{macro1}
\begin{equation}\label{twononinterpulses}
\begin{cases}
\ds \partial_\tau A_{1,1}-\nabla_\vth\Omega(\vth_1){\cdot}\nabla_y A_{1,1}=0,
\\[2mm]
\ds \partial_\tau A_{1,2}-\nabla_\vth\Omega(\vth_2){\cdot}\nabla_y A_{1,2}=0,
\end{cases}
\end{equation}
and the functions $A_{2,(j,j)}$ ($j=1,2$), $A_{2,(1,\pm2)}$, $A_{2,(1,-1)}$ 
by \eqref{macro2}
\begin{equation*}
A_{2,(j,j)}=\frac{c_{(j,j)}}{\delta_{(j,j)}}A_{1,j}^2,
\quad
A_{2,(1,\pm2)}=\frac{2c_{(1,\pm2)}}{\delta_{(1,\pm2)}}A_{1,1}A_{1,\pm2},
\quad
A_{2,(1,-1)}
=
\frac{-2b_2}{b_1}(|A_{1,1}|^2+|A_{1,2}|^2)
\end{equation*}
with $c_{(p,q)}$ given by \eqref{c2}. 

Note, that here all microscopic patterns within the $\ve^2$-term of 
\eqref{nu2N2} are mutually different, except when 
either $(3\vth_1,3\om_1)=(\vth_2,\om_2)$ or $(\vth_1,\om_1)=(3\vth_2,3\om_2)$,
where we obtain only one respresentant for 
either $\{(1,1),(-1,2)\}$ or $\{(2,2),(1,-2)\}$, respectively,
and the corresponding amplitudes are given by either 
$A_{2,(1,1)}+\ol{A_{2,(1,-2)}}$ or $A_{2,(2,2)}+A_{2,(1,-2)}$
with the values from above.

Hence, in the present case the two pulses do not interact with each other 
in leading order. 
Their first-order amplitudes pass through each other travelling 
with the respective group-velocities of the pulses they modulate,
i.e., the corresponding transport equations are uncoupled.
Thus, ignoring one of the two amplitudes 
(i.e. assuming it vanishes identically in space and time)
the other remains unchanged, and $X_2^{A,\ve}$ is the same as for a 
single $2$-closed pulse, see Sec.\ \ref{singleclosed2}.\\

\noindent{\bf Case 2: Interacting pulses.}\\[2mm]
%
In the case $(2\vth_1,2\om_1)=(\vth_2,\om_2)$
we have $\delta_{(1,1)}=\delta_2=\delta_{(1,-2)}=\delta_{-1}=0$,
the closedness condition \eqref{nu2N2clcond} 
reduces to the nonresonance conditions
\begin{equation}
\delta_{(2,2)}=\Om^2(4\vth_1)-16\om_1^2\neq 0,
\qquad
\delta_{(1,2)}=\Om^2(3\vth_1)-9\om_1^2\neq 0,
\label{nu2N2interclcond}
\end{equation}
and the equations \eqref{macro1} read
\begin{equation}
\begin{cases}
\ds \  \ \! \partial_\tau A_{1,1}
-\nabla_\vth\Omega(\vth_1){\cdot}\nabla_y A_{1,1}
=-\i\frac{
\ol{c_{(1,1)}}
}{\Omega(\vth_1)}A_{1,2}\ol{A_{1,1}},
\\[5mm]
\ds \partial_\tau A_{1,2}
-\nabla_\vth\Omega(2\vth_1){\cdot}\nabla_y A_{1,2}
=-\i\frac{c_{(1,1)}}{4\Omega(\vth_1)}A_{1,1}^2
\end{cases}
\label{nu2selfamp}
\end{equation}
with $c_{(1,1)}$ given by \eqref{c11}.
Since $\at_2\setminus\calN=\{\pm(2,2),\pm(1,2),(0,0)\}$, we obtain by 
\eqref{macro2} the functions $A_{2,(2,2)}$, $A_{2,(1,2)}$, $A_{2,(1,-1)}$ 
as in Case 1,
whereas the functions $A_{2,(1,1)}$, $A_{2,(1,-2)}$ are included in 
$A_{2,2}$, $\ol{A_{2,1}}$, respectively, which remain undetermined. 
(Compare also the values of the former functions in Case 1 with the right-hand
sides of \eqref{nu2selfamp}.)

Here,
the evolution equations for the amplitudes are coupled,
in an asymmetric way that reflects that the pulse $(\vth_2,\om_2)$
is generated by self-interaction of $(\vth_1,\om_1)$.
In particular, setting identically $A_{1,1}=0$, \eqref{nu2selfamp} becomes a
homogeneous transport equation for $A_{1,2}$,
whereas, setting $A_{1,2}=0$, \eqref{nu2selfamp} yields $A_{1,1}=0$.
The latter is exactly the situation in a non-$2$-closed system of a single
pulse, see Sec.\ \ref{singlenonclosed2}.
In the contrary, we can describe the dynamics of such a self-interacting pulse
by \eqref{nu2selfamp}, by taking into account via the ansatz \eqref{nu2N2} 
also the amplitude $A_{1,2}$ of the generated pulse, 
provided of course the system is $2$-closed.
(Note in particular how even for $A_{1,2}(0,\cdot)=0$ a 
non-zero  $A_{1,2}$
is generated by  \eqref{nu2selfamp}, when $A_{1,1}(0,\cdot)\neq0$.)
The situation of non-$2$-closed systems of two pulses is discussed 
in the following.
%
\subsubsection{Non-2-closed systems of two pulses}
\label{twopulsesnonclosedN2}
\noindent{\bf Case 1: Non-interacting pulses.}\\[2mm]
The closedness condition \eqref{nu2N2clcond} is violated
if one of the pulses generates via self-interaction a third, 
$(\vth_3,\om_3)=(2\vth_j,2\om_j)$, $j=1,2$, 
or if the two pulses interact to generate a third, 
$(\vth_3,\om_3)=(\vth_1\pm\vth_2,\om_1\pm\om_2)$.
In the present case neither of these pulses equals 
$\pm(\vth_j,\om_j)$, $j=1,2$. 
Of course, more than one of the four 
(or three, if $(3\vth_j,3\om_j)=(\vth_i,\om_i)$, $j\neq i$) 
corresponding nonresonance conditions \eqref{nu2N2clcond} can be violated,
and moreover, even if only one new pulse arises it is not at all clear
that the three pulses form a $2$-closed set.

Hence, we restrict our discussion
to two typical cases where only one of the four 
conditions \eqref{nu2N2clcond} is violated.
In the first case, we assume $\delta_{(2,2)}=0$.
Then according to the equation for $A_{2,2}$ 
in Sec.\ \ref{twopulsesclosedN2}, Case 1,
it follows $A_{1,2}=0$ 
(if $c_{(2,2)}\neq0$, e.g., when $b_2\neq0$ in \eqref{c2}).
Then, as discussed 
there
(and setting $A_{2,(2,2)}=0$), 
we can still describe the dynamics of $A_{1,2}$ as in 
Sec.\ \ref{singleclosed2}.
However, we are not able to determine any non-trivial amplitude $A_{1,2}$ 
for the self-interacting pulse $\pm(\vth_2,\om_2)$ 
(for instance by assuming initially $A_{1,2}(0,\cdot)\neq 0$).
A first step to remedy this is to include the pulse 
$(\vth_3,\om_3)=(2\vth_2,2\om_2)$ 
into the set of considered pulses, 
see Sec.\ \ref{threepulsesclosedN2}, Case \ref{ca3}.

In the second case we assume $\delta_{(1,2)}=0$, which implies 
$A_{1,1}A_{1,2}=0$, 
see the equation for $A_{2,(1,\pm2)}$ in Sec.\ \ref{twopulsesclosedN2}, C.\ 1.
%
Hence, $A_{1,1}$ and $A_{1,2}$ should be for all times disjointly supported.
Except for $A_{1,1}=0$ or $A_{1,2}=0$ identically, leading again to
the case of a single pulse, this violates the essence of interaction of the 
corresponding pulses, which thus can not be explained.
(Apart from restricting the choice of allowed initial data for the amplitudes, 
this fails to predict their evolution e.g.\ if they move towards each other.)
%
%
The correct set-up 
is 
once 
again to take into account also 
the generated pulse, 
see Sec.\ \ref{threepulsesclosedN2}, Case \ref{case3wi}.\\

\noindent{\bf Case 2: Interacting pulses.}\\[2mm]
%
The closedness condition \eqref{nu2N2interclcond} is violated if 
$\delta_{(2,2)}=0$ 
or $\delta_{(1,2)}=0$,
generating a third pulse, $(\vth_3,\om_3)=(4\vth_1,4\om_1)$ 
or $(\vth_3,\om_3)=(3\vth_1,3\om_1)$, different from 
$\pm(\vth_j,\om_j)$, $j=1,2$. 
The respective $2$-closed systems of three pulses are discussed in 
Sec.\ \ref{threepulsesclosedN2}, Cases \ref{ca4}, \ref{ca5}.
Here, $\delta_{(2,2)}=0$ yields as above $A_{1,2}=0$,  
but moreover, $A_{1,1}=0$ by \eqref{nu2selfamp}.
%
Thus, noting that $(\vth_2,\om_2)=(2\vth_1,2\om_1)$ is generated via
self-interaction, we observe the same phenomenon mentioned already in 
Sec.\ \ref{singlenonclosed2} and Sec.\ \ref{twopulsesclosedN2}, Case 2:
\begin{center}
\emph{if we ignore a generated pulse (by assuming it has an identically 
vanishing amplitude), then we have to ignore also (at least) one of 
its generators.}
\end{center}
In the second case, $\delta_{(1,2)}=0$, we obtain 
$A_{1,1}A_{1,2}=0$ 
as in Case 1.
Apart from the implications discussed there, 
here the macroscopic system \eqref{nu2selfamp} shows clearly 
an inherent direction in the generation of pulses:
while $A_{1,2}=0$ yields $A_{1,1}=0$, 
assuming $A_{1,1}=0$ we still can describe the dynamics of $A_{1,2}$.
Thus, 
\begin{center}
\emph{generating pulses can be ignored, while generated ones can not.} 
\end{center}
%
%
%
Hence, in order to obtain relevant macroscopic dynamics for a system of 
modulated pulses  
we have to take into account also the pulses generated by them,
i.e., 
to require that the set is closed with respect to interactions,
motivating
Definition \ref{defclosure}.

Concerning the present case, when two pulses generate a third, new one, 
the appropriate setting for describing their dynamics is given by considering
the set of the three pulses, provided of course it is closed.
We do this in the following.
%
\subsection{Three pulses}
\label{threepulses}
We consider the case $\nu=3$ of three different pulses 
$(\vth_j,\om_j)\neq(\vth_i,\om_i)$ for $i\neq j$ and $i,j=1,2,3$,
with $\om_j=\Om(\vth_j)>0$.
The second-order approximation \eqref{XNA} 
is given by
\begin{multline}\label{nu3N2}
X_2^{A,\ve}
=\ve\sum_{j=1}^3 A_{1,j}\boe_j
+\ve^2\Big(
\sum_{j=1}^3 \big(A_{2,j}\boe_j+A_{2,(j,j)}\boe_j^2\big)
+\sum_{1\le j<i\le 3}
\big(A_{2,(j,i)}\boe_j\boe_i+A_{2,(j,-i)}\boe_j\boe_{-i}\big)
\\+{\frac12} A_{2,(1,-1)}
\Big)+\cc,
\end{multline}
with $\boe_j(t,\g)=\e^{\i(\om_j t+\vth_j\cdot\g)}$, $j=\pm1,\pm2,\pm3$, 
where $\om_{-j}=-\om_{j}$, $\vth_{-j}=-\vth_{j}$.
Here, $(1,\pm2)$, $(1,\pm3)$ and $(2,\pm3)$ represent also $(\pm2,1)$,
$(\pm3,1)$ and $(\pm3,2)$, respectively, 
and $(1,-1)$ represents $\pm(j,-j)$, $j=1,2,3$.
Else, we start from the assumption that 
$\{j, (j,j): j=1,2,3\}\cup\{(1,2),(1,-2),(1,3),(1,-3),(2,3),(2,-3)\}$ 
is a set of (different) representants.

In Section \ref{threepulsesclosedN2} we give the explicit 
evolution equations for the first-order amplitudes $A_{1,j}$, $j=1,2,3$, 
and the functions $A_{2,\whj_m}$, $\whj_m\in\at_2\setminus\calN$
of the approximation \eqref{nu3N2} for a set which is closed under
interactions up to order $N=2$.
The equations depend on the presence, the number and the kind of
interactions within the set. 

In Section 
\ref{432} we consider a non-$2$-closed set for a three-wave-interaction
and give the equations for the first-order amplitudes of a 
$2$-closed system 
of four pulses with two three-wave-interactions.
%
 Finally, in Section
 \ref{433} we determine a third-order approximation $X_3^{A,\ve}$ 
 for the $3$-closed system of the three-wave-interaction 
of  Sec.\ \ref{432}, Case \ref{case3wi}.
%
\subsubsection{2-closed systems of three pulses}
\label{threepulsesclosedN2}
According to Definition \ref{defclosure}, 
the set of three different pulses $\{\pm(\vth_j,\om_j):\ j=1,2,3\}$ 
is closed under interactions up to order $2$ if 
\begin{center}
\emph{the pairs $(2\vth_j,2\om_j)$, $(\vth_j{\pm}\vth_i,\om_j{\pm}\om_i)$,
$j\neq i$, ($j,i=1,2,3$) 
do not characterize pulses,\\
except if 
$(\vth_j{+}\vth_i,\om_j{+}\om_i)=(\vth_k,\om_k)$, $k\neq j,i$.}
\end{center}
Since the latter exceptions correspond to interactions among the three pulses,
we can classify the possible settings according to the number of interactions
and the pulses involved, up to relabeling, as follows: 
\begin{enumerate}
\item\label{ca1}
No interactions.
\item\label{case3wi}
Only a three-wave-interaction: 
$(\vth_1+\vth_2,\om_1+\om_2)=(\vth_3,\om_3)$.
\item\label{ca3}
Only one self-interaction: 
$(2\vth_1,2\om_1)=(\vth_2,\om_2)$.
\item\label{ca4} 
Two self-interactions: 
$(2\vth_1,2\om_1)=(\vth_2,\om_2)$, 
$(2\vth_2,2\om_2)=(4\vth_1,4\om_1)=(\vth_3,\om_3)$.
\item\label{ca5}
One self-interaction and one three-wave-interaction:
$(2\vth_1,2\om_1)=(\vth_2,\om_2)$,\\ 
$(\vth_1+\vth_2,\om_1+\om_2)=(3\vth_1,3\om_1)=(\vth_3,\om_3)$.
\end{enumerate}

We assume that the nonresonance conditions given in the following are 
independent. 
It is possible that for special choices of pulses they can be reduced, cf.\
the example of Sec.\ \ref{twopulsesclosedN2}, Case 1.
However, this does not affect the first-order amplitude equations, but only
the coefficients of 'non-pulses'. 
Hence, the following classification remains unaffected.\\ 

\noindent{\bf Case \ref{ca1}: No interactions.}\\[2mm]
The closedness condition is equivalent to the nonresonance conditions 
\begin{equation*}\label{nu3N2ca1clcond}
\delta_{(j,j)}\neq 0,\quad j=1,2,3,\qquad
\delta_{(1,\pm2)}\neq 0,\qquad
\delta_{(1,\pm3)}\neq 0,\qquad
\delta_{(2,\pm3)}\neq 0,
\end{equation*}
with $\delta_{(j,\pm i)}=\Om^2(\vth_j{\pm}\vth_i)-(\om_j{\pm}\om_i)^2$
for $j,i=1,2,3$,
and \eqref{macro1} yields
\begin{equation*}
\begin{cases}
\ds 
\partial_\tau A_{1,1}-\nabla_\vth\Omega(\vth_1){\cdot}\nabla_y A_{1,1}
=0,
\\[2mm]
\ds \partial_\tau A_{1,2}-\nabla_\vth\Omega(\vth_2){\cdot}\nabla_y A_{1,2}
=0,
\\[2mm]
\ds \partial_\tau A_{1,3}-\nabla_\vth\Omega(\vth_3){\cdot}\nabla_y A_{1,3}
=0.
\end{cases}
\end{equation*}
By \eqref{macro2} we obtain ($j,i=1,2,3$)
\begin{align*}
A_{2,(j,j)}=\frac{c_{(j,j)}}{\delta_{(j,j)}}A_{1,j}^2,
\quad
A_{2,(j,\pm i)}=\frac{2c_{(j,\pm i)}}{\delta_{(j,\pm i)}}
A_{1,j}A_{1,\pm i}\ \ (j<i),\quad
A_{2,(1,-1)}=
\frac{-2b_2}{b_1}\sum_{j=1}^3|A_{1,j}|^2.
\end{align*}
Since the pulses do not interact, 
the amplitudes $A_j$ are transported independently of each other by the 
respective group velocities $\nabla_\vth\Omega(\vth_j)$.
If their trajectories intersect, they simply pass through each other.
Since the equations are uncoupled, neglecting one or two amplitudes 
by setting identically $A_{1,j}=0$, 
yields the corresponding systems for two or one pulse, 
see Sec.\ \ref{twopulsesclosedN2}, Case 1, and Sec.\ \ref{singleclosed2}.\\

\noindent{\bf Case \ref{case3wi}: Only a three-wave-interaction.}\\[2mm]
We assume that the resonance condition 
\begin{equation}\label{ca2rescond}
(\vth_1{+}\vth_2,\om_1{+}\om_2)=(\vth_3,\om_3)
\end{equation}
holds, and that self-interactions are absent. 
The closedness condition is equivalent to the nonresonance conditions
\begin{equation}\label{nu3N2ca2clcond}
\delta_{(j,j)}\neq 0,\quad j=1,2,3,\qquad
\delta_{(1,3)}\neq 0,\qquad
\delta_{(1,-2)}\neq 0,\qquad
\delta_{(2,3)}\neq 0,
\end{equation}
while $\delta_{(1,2)}=\delta_{3}=\delta_{(-1,3)}=\delta_{2}=\delta_{(-2,3)}
=\delta_{1}=0$.
Then, 
assuming that the nonresonance conditions \eqref{nu3N2ca2clcond}
are mutually independent,
the second-order approximation \eqref{nu3N2} becomes
\begin{multline}\label{nu3N2inter}
X_2^{A,\ve}
=\ve\sum_{j=1}^3 A_{1,j}\boe_j
+\ve^2\Big(\frac12 A_{2,(1,-1)}+\sum_{j=1}^3 \Big(A_{2,j}\boe_j+A_{2,(j,j)}
\boe_j^2
\Big)
\\+A_{2,(1,3)}
\boe_1\boe_3
+A_{2,(1,-2)}
\boe_1\boe_{-2}
+A_{2,(2,3)}
\boe_2\boe_3
\Big)+\cc
\end{multline}
By \eqref{macro1}, the equations for the amplitudes $A_{1,j}$, $j=1,2,3$, read
\begin{equation}\label{3wi}
\begin{cases}
\ds \partial_\tau A_{1,1}-\nabla_\vth\Omega(\vth_1){\cdot}\nabla_y A_{1,1}
=-\i\frac{\ol{c_{(1,2)}}}{\Omega(\vth_1)}A_{1,3}\ol{A_{1,2}},
\\[5mm]
\ds \partial_\tau A_{1,2}-\nabla_\vth\Omega(\vth_2){\cdot}\nabla_y A_{1,2}
=-\i \frac{\ol{c_{(1,2)}}}{\Omega(\vth_2)}A_{1,3}\ol{A_{1,1}},
\\[5mm]
\ds \partial_\tau A_{1,3}-\nabla_\vth\Omega(\vth_3){\cdot}\nabla_y A_{1,3}
=-\i \frac{c_{(1,2)}}{\Omega(\vth_3)}A_{1,1}A_{1,2}
\end{cases}
\end{equation}
with $\vth_3=\vth_1+\vth_2$, 
$\Omega(\vth_3)=\Omega(\vth_1)+\Omega(\vth_2)$ and
\begin{align}
c_{(1,2)}
&=
-4\i\sum_{\alpha\in\Gamma}a_{2,\alpha}
\sin\left(\frac{\vth_1}2{\cdot}\alpha\right)
\sin\left(\frac{\vth_2}2{\cdot}\alpha\right)
\sin\left(\frac{\vth_3}2{\cdot}\alpha\right)
-b_2.
\label{c12}
\end{align}
%
By \eqref{macro2} 
and \eqref{nu3N2ca2clcond}
we obtain 
as in Case \ref{ca1}
 \begin{align}\label{3wimixed}
& A_{2,(j,j)}=\frac{c_{(j,j)}}{\delta_{(j,j)}}A_{1,j}^2,\quad j=1,2,3,
\qquad
 A_{2,(1,-1)}=\frac{-2b_2}{b_1}\sum_{j=1}^3|A_{1,j}|^2,
\notag\\&
A_{2,(1,3)}=\frac{2c_{(1,3)}}{\delta_{(1,3)}}A_{1,1}A_{1,3},
\quad
A_{2,(1,-2)}=\frac{2c_{(1,-2)}}{\delta_{(1,-2)}}A_{1,1}
\ol{A_{1,2}},
\quad
A_{2,(2,3)}=\frac{2c_{(2,3)}}{\delta_{(2,3)}}A_{1,2}A_{1,3}
 \end{align}
with $c_{(p,q)}$ given by \eqref{c2},
while 
the functions $A_{2,(1,2)}$, $A_{2,(1,-3)}$,  
and $A_{2,(2,-3)}$  are included in 
$A_{2,3}$, $\ol{A_{2,2}}$, and $\ol{A_{2,1}}$, respectively, 
since $(1,2)=3,(1,-3)=-2,(2,-3)=-1\in\calN$,
and thus remain undetermined by \eqref{macro1}, \eqref{macro2}.

The system \eqref{3wi} of three nonlinearly coupled transport equations, 
the \emph{three-wave interaction equation} 
(cf.\ \cite{BS90,Kau80,Kea99,KRB79,SW03}), 
is the generic system describing the macroscopic dynamics of 
{\it three different} modulated pulses, 
interacting {\it all} with each other, 
each being generated by the other two:
no subsystem decouples and the nonlinearity is symmetric 
with respect to the amplitudes. 
%

Ignoring any of the involved pulses by setting, 
e.g., $A_{1,3}=0$,
leads to 
the system 
\begin{align*}
\partial_\tau A_{1,1}-\nabla_\vth\Omega(\vth_1){\cdot}\nabla_y A_{1,1}
=0,\qquad
\partial_\tau A_{1,2}-\nabla_\vth\Omega(\vth_2){\cdot}\nabla_y A_{1,2}
=0,\qquad
A_{1,1}A_{1,2}=0.
\end{align*}
This is the case of the non-$2$-closed system of two pulses which 
generate via interaction a third, discussed in 
Sec.\ \ref{twopulsesnonclosedN2}, Case 1,
which, as we saw there, 
prohibits a complete description of the dynamics of the two generating pulses,
due to the condition $A_{1,1}A_{1,2}=0$. 
To obtain the correct dynamics one has to consider \eqref{3wi}, provided of
course the system of the three pulses is closed.\\

\noindent{\bf Case \ref{ca3}: Only one self-interaction.}\\[2mm]
We assume that the resonance condition $(2\vth_1,2\om_1)=(\vth_2,\om_2)$ 
holds, and that no other interactions are present.
This leads to $\delta_{(1,1)}=\delta_{2}=\delta_{(1,-2)}=\delta_{-1}=0$, 
and the closedness condition is given by
the nonresonance conditions
\begin{equation*}
\delta_{(j,j)}\neq 0,\quad j=2,3,\qquad
\delta_{(1,2)}\neq 0,\qquad
\delta_{(1,\pm3)}\neq 0,\qquad
\delta_{(2,\pm3)}\neq 0.
\end{equation*}
The equations \eqref{macro1} yield
\begin{equation*}
\begin{cases}
\ds\  \ \! 
\partial_\tau A_{1,1}-\nabla_\vth\Omega(\vth_1){\cdot}\nabla_y A_{1,1}
=-\i\frac{\ol{c_{(1,1)}}}{\Omega(\vth_1)}A_{1,2}\ol{A_{1,1}},
\\[5mm]
\ds \partial_\tau A_{1,2}-\nabla_\vth\Omega(2\vth_1){\cdot}\nabla_y A_{1,2}
=-\i\frac{c_{(1,1)}}{4\Omega(\vth_1)}A_{1,1}^2,
\\[5mm]
\ds\  \ \! 
\partial_\tau A_{1,3}-\nabla_\vth\Omega(\vth_3){\cdot}\nabla_y A_{1,3}
=0
\end{cases}
\end{equation*}
with $c_{(1,1)}$ given by \eqref{c11}.
By \eqref{macro2} we obtain the same equations for 
$A_{2,(j,j)}$, $j=2,3$,  $A_{2,(1,2)}$, $A_{2,(i,\pm3)}$, $i=1,2$, 
$A_{2,(1,-1)}$, as in Case \ref{ca1}, 
while $A_{2,(1,1)}$, $A_{2,(1,-2)}$ are included in $A_{2,2}$, $\ol{A_{2,1}}$, 
respectively,
and thus remain undetermined.

Here, 
in contrast to
\eqref{3wi},
the subsystem for the amplitudes $A_{1,1}$, $A_{1,2}$,
given already by \eqref{nu2selfamp},
decouples from the transport equation for $A_{1,3}$, 
since the set $\{\pm(\vth_j,\om_j): j=1,2\}$ 
forms itself a $2$-closed system,
which does not interact with the third pulse.
The amplitude of the latter travels through the other two unperturbed.
Up to relabeling, the full system is the closure of the two non-interacting 
pulses with $\delta_{(2,2)}=0$ of Sec.\ \ref{twopulsesnonclosedN2}, Case 1.


In this context, we would like to mention the trivial fact that 
although the resonance condition $(2\vth_1,2\om_1)=(\vth_3,\om_3)$ 
is a special case of $(\vth_1+\vth_2,\om_1+\om_2)=(\vth_3,\om_3)$, 
where $(\vth_1,\om_1)=(\vth_2,\om_2)$,
%
the correct dynamics of the first-order amplitudes 
$A_{1,1}+A_{1,2}$ and $A_{1,3}$ 
of the specified approximation $X_2^{A,\ve}$, see \eqref{nu3N2},
cannot be obtained directly from \eqref{3wi}.
Indeed, in this case the latter gives
 \begin{align*}
 \partial_\tau \big(A_{1,1}+A_{1,2}\big)
 -\nabla_\vth\Omega(\vth_1){\cdot}\nabla_y\big(A_{1,1}+A_{1,2}\big)
 &=-\i\frac{\ol{c_{(1,1)}}}{\Omega(\vth_1)}A_{1,3}
 \ol{(A_{1,1}+A_{1,2})},
 \\
 \partial_\tau A_{1,3}-\nabla_\vth\Omega(2\vth_1){\cdot}\nabla_y A_{1,3}
 &=-\i \frac{c_{(1,1)}}{
 4\Omega(\vth_1)
 }\left(2A_{1,1}A_{1,2}
 +\Big[A_{1,1}^2+A_{1,2}^2\Big]\right)
 \end{align*}
{\it without} the terms in square brackets.
However, the latter are included in the correct macroscopic equations for 
the amplitudes $A_{1,1}+A_{1,2}$ and $A_{1,3}$ of two pulses with 
$(2\vth_1,2\om_1)=(\vth_3,\om_3)$ forming a $2$-closed system,
cf.\  \eqref{nu2selfamp}, i.e., the subsystem from above.
The reason for this inconsistency lies in the different form of the 
underlying multiscale  ansatz'es \eqref{nu3N2} vs.\ \eqref{nu2N2}.
Naturally, 
the correct ansatz for a given (closed) system of pulses, 
is prescribed by the {\it different} pulses among them.\\  

\noindent{\bf Case \ref{ca4}: Two self-interactions.}\\[2mm]
We assume $(2\vth_1,2\om_1)=(\vth_2,\om_2)$ 
and $(2\vth_2,2\om_2)=(4\vth_1,4\om_1)=(\vth_3,\om_3)$.
Hence, $\delta_{(1,1)}=\delta_{2}=\delta_{(2,2)}=\delta_{3}=\delta_{(1,-2)}
=\delta_{-1}=\delta_{(2,-3)}=\delta_{-2}=0$, 
and the closedness 
of the system 
is guaranteed by the nonresonance conditions
\begin{equation*}
\delta_{(3,3)}\neq 0,\qquad
\delta_{(1,2)}=\delta_{(1,-3)}\neq 0,\qquad
\delta_{(1,3)}\neq 0,\qquad
\delta_{(2,3)}\neq 0.
\end{equation*}
The equations \eqref{macro1} yield
\begin{equation*}
\begin{cases}
\ds\ \ \! 
\partial_\tau A_{1,1}-\nabla_\vth\Omega(\vth_1){\cdot}\nabla_y A_{1,1}
=-\i\frac{\ol{c_{(1,1)}}}{\Omega(\vth_1)}\ol{A_{1,1}}A_{1,2},
\\[5mm]
\ds \partial_\tau A_{1,2}-\nabla_\vth\Omega(2\vth_1){\cdot}\nabla_y A_{1,2}
=-\i\frac1{4\Omega(\vth_1)}\big(c_{(1,1)}A_{1,1}^2
+2\ol{c_{(2,2)}}\ol{A_{1,2}}A_{1,3}\big),
\\[5mm]
\ds \partial_\tau A_{1,3}-\nabla_\vth\Omega(4\vth_1){\cdot}\nabla_y A_{1,3}
=-\i\frac{c_{(2,2)}}{8\Omega(\vth_1)}A_{1,2}^2
\end{cases}
\end{equation*}
with $c_{(1,1)}$ as in \eqref{c11} and 
 \begin{align*}
  c_{(2,2)}&=-4\i\sum_{\alpha\in\Gamma}a_{2,\alpha}
  \sin^2\left(
\vth_1
{\cdot}\alpha\right)
  \sin\left(
2\vth_1
{\cdot}\alpha\right)
  -b_2.
 \end{align*}
By \eqref{macro2} we obtain 
$A_{2,(j,3)}$, $j=1,2,3$, $A_{2,(1,-1)}$ as in Case \ref{ca1},
and 
\begin{align*}
%
%
A_{2,(1,2)}
=\frac2{\delta_{(1,2)}}
\big(c_{(1,2)}A_{1,1}A_{1,2}+c_{(-1,3)}
\ol{A_{1,1}}
A_{1,3}\big).
\end{align*}
%
Note that here $(1,2)$ represents also $(-1,3)$, since 
$\boe_1\boe_2=\boe_3\boe_{-1}(=\boe_1^3)$.
Hence, $\ol{A_{2,(1,-3)}}$ is included in $A_{2,(1,2)}$.
The functions $A_{2,(1,1)}$, $A_{2,(2,2)}$, 
$A_{2,(1,-2)}$, $A_{2,(2,-3)}$ are included in 
$A_{2,2}$, $A_{2,3}$, 
$\ol{A_{2,1}}$, $\ol{A_{2,2}}$, respectively,
and hence remain undetermined.

In the present case the third pulse is generated by self-interaction of the
second, which in turn is generated by self-interaction of the first. 
This is reflected in the above equations, 
where we observe that 
(a) $A_{1,3}=0$ yields $A_{1,2}=A_{1,1}=0$; 
(b) $A_{1,2}=0$ implies $A_{1,1}=0$, 
while $A_{1,3}$ evolves according to an uncoupled transport equation;
and (c) $A_{1,1}=0$ yields a coupled system for the dynamics 
of $A_{1,2}$ and $A_{1,3}$
of the form \eqref{nu2selfamp}, 
%
%
cf.\ the discussion in Sec.\ \ref{twopulsesnonclosedN2}, Case 2.
(In particular, the above system is the closure of the two interacting pulses
with $\delta_{(2,2)}=0$ discussed there.)\\
%

\noindent{\bf Case \ref{ca5}: One self-interaction and one 
three-wave-interaction.}\\[2mm]
We assume $(2\vth_1,2\om_1)=(\vth_2,\om_2)$ 
and $(\vth_1{+}\vth_2,\om_1{+}\om_2)=(3\vth_1,3\om_1)=(\vth_3,\om_3)$.
This yields the resonance conditions
$\delta_{(1,1)}=\delta_{2}=\delta_{(1,2)}=\delta_{3}
=\delta_{(1,-2)}=\delta_{-1}=\delta_{(1,-3)}=\delta_{-2}=
\delta_{(2,-3)}=\delta_{-1}=0$,
and the closedness condition is equivalent to the nonresonance conditions 
\begin{equation*}
\delta_{(2,2)}=\delta_{(1,3)}\neq 0,\qquad
\delta_{(3,3)}\neq 0,\qquad
\delta_{(2,3)}\neq 0.
\end{equation*}
Then, the equations \eqref{macro1} yield
\begin{equation*}
\begin{cases}
\ds\  \ \! 
\partial_\tau A_{1,1}-\nabla_\vth\Omega(\vth_1){\cdot}\nabla_y A_{1,1}
=-\i\frac1{\Omega(\vth_1)}
\left(\ol{c_{(1,2)}}A_{1,3}\ol{A_{1,2}}+\ol{c_{(1,1)}}\ol{A_{1,1}}A_{1,2}
\right),
\\[5mm]
\ds \partial_\tau A_{1,2}-\nabla_\vth\Omega(2\vth_1){\cdot}\nabla_y A_{1,2}
=-\i\frac1{2\Omega(\vth_1)}
\left(\ol{c_{(1,2)}}A_{1,3}\ol{A_{1,1}}+\frac{c_{(1,1)}}2A_{1,1}^2\right),
\\[5mm]
\ds \partial_\tau A_{1,3}-\nabla_\vth\Omega(3\vth_1){\cdot}\nabla_y A_{1,3}
=-\i\frac{c_{(1,2)}}{3\Omega(\vth_1)}A_{1,1}A_{1,2}
\end{cases}
\end{equation*}
with $c_{(1,1)}$ and $c_{(1,2)}$ given by \eqref{c11} and  
\eqref{c12} with $\vth_2=2\vth_1$ and $\vth_3=3\vth_1$, respectively.
By \eqref{macro2} we obtain $A_{2,(3,3)}$, $A_{2,(2,3)}$, $A_{2,(1,-1)}$ 
as in Case \ref{ca1}, and
\begin{align*}
A_{2,(2,2)}
=\frac1{\delta_{(2,2)}}
\big(c_{(2,2)}A_{1,2}^2+2c_{(1,3)}A_{1,1}A_{1,3}\big).
\end{align*}
%
Here, $(2,2)$ represents also $(1,3)$, 
since $\boe_2^2=\boe_1\boe_3(=\boe_1^4)$.
Hence, $A_{2,(1,3)}$ is included in $A_{2,(2,2)}$.
The functions 
$A_{2,(1,1)}$, $A_{2,(1,2)}$, $A_{2,(1,-2)}$, $A_{2,(1,-3)}$, $A_{2,(2,-3)}$ 
are included in 
$A_{2,2}$, $A_{2,3}$, $\ol{A_{2,1}}$, $\ol{A_{2,2}}$, $\ol{A_{2,1}}$,
respectively, 
and hence remain undetermined.
Note, that the first terms on the right hand sides of the 
amplitude
equations are the same as in \eqref{3wi}. 
However, due to the additional self-interaction, 
we obtain also the corresponding second terms.

Here, 
assuming (a) $A_{1,3}=0$, yields the system \eqref{nu2selfamp} under the
condition  $A_{1,2}A_{1,1}=0$, cf.\  
Sec.\ \ref{twopulsesnonclosedN2}, Case 2 (for $\delta_{(1,2)}=0$).
If $A_{1,2}=0$ we further get $A_{1,1}=0$, while if $A_{1,1}=0$ 
the system reduces to a transport equation for $A_{1,2}$.
Assuming (b) $A_{1,2}=0$, we obtain
\begin{equation*}
\partial_\tau A_{1,1}-\nabla_\vth\Omega(\vth_1){\cdot}\nabla_y A_{1,1}=0,
\ \  
\partial_\tau A_{1,3}-\nabla_\vth\Omega(\vth_3){\cdot}\nabla_y A_{1,3}=0,
\ \
A_{1,1}^2=\frac{-2
\ol{c_{(1,2)}}
}{c_{(1,1)}}\ol{A_{1,1}}A_{1,3}.
\end{equation*}
If $A_{1,1}=0$ the evolution of $A_{1,3}$ can still be described,
while $A_{1,3}=0$ implies $A_{1,1}=0$.
Even for $A_{1,3}\neq0$ these equations yield  
$0=\ol{A_{1,1}}\big(\nabla_\vth\Omega(\vth_3)-\nabla_\vth\Omega(\vth_1)\big)
{\cdot}\nabla_y A_{1,3}$
on the support of  $A_{1,3}$, which in general implies $A_{1,1}=0$.
Finally, assuming (c) $A_{1,1}=0$, we obtain the system 
\eqref{twononinterpulses} and $A_{1,3}\ol{A_{1,2}}=0$, 
which in general imply that one amplitude vanishes
identically allowing to describe the dynamics of the other.
Considering the succesion in the generation of pulses in the present case, 
these observations confirm the principles postulated in 
Sec.\ \ref{twopulsesnonclosedN2}.

%
\subsubsection{Non-2-closed system of a three-wave-interaction}
\label{432}
%
%
%
We consider a system of three different pulses 
$\{\pm(\vth_j,\om_j): j=1,2,3\}$ with $\om_j=\Om(\vth_j)>0$,
which satisfies the resonance condition \eqref{ca2rescond},
$(\vth_1+\vth_2,\om_1+\om_2)=(\vth_3,\om_3)$, 
of Case \ref{case3wi} in Sec.\ \ref{threepulsesclosedN2},
but which is not closed under interactions up to order $2$.
In particular we assume that the nonresonance condition $\delta_{(2,3)}\neq 0$
of \eqref{nu3N2ca2clcond} is violated, while the others are still satisfied.
Thus, the pulse $(\vth_2+\vth_3,\om_2+\om_3)=(\vth_1+2\vth_2,\om_1+2\om_2)
=(\vth_4,\om_4)$ is generated by the interaction of the former two,
and the set $\{\pm(\vth_j,\om_j): j=1,\ldots,4\}$ consists 
of four different pulses.

As we saw in the previous sections, the effect of non-$2$-closedness on a 
system of pulses results from ignoring the amplitudes of the generated pulses.
%
Thus, assuming that the set of the four pulses is $2$-closed,
we determine next the evolution equations for its first-order
amplitudes, and discuss subsequently the effects of 
ignoring the amplitude of $\pm(\vth_4,\om_4)$.

The second-order approximation \eqref{XNA} for $\nu=4$ different pulses reads
 \begin{multline*}
X_2^{A,\ve}
=\ve\sum_{j=1}^4 A_{1,j}\boe_j
+\ve^2\Big(
\sum_{j=1}^4 \big(A_{2,j}\boe_j+A_{2,(j,j)}\boe_j^2\big)
\\
+\sum_{1\le j<i\le 4}
\big(A_{2,(j,i)}\boe_j\boe_i+A_{2,(j,-i)}\boe_j\boe_{-i}\big)
+{\frac12} A_{2,(1,-1)}\Big)+\cc
\end{multline*}
with $\boe_j(t,\g)=\e^{\i(\om_j t+\vth_j\cdot\g)}$, $j=1,\ldots,4$, 
and when all appearing indices are representants.
By the assumed interactions we have the resonance conditions
\begin{equation*}
\delta_{(1,2)}
=\delta_{(1,-3)}
=\delta_{(2,-3)}
=\delta_{(2,3)}
=\delta_{(2,-4)}
=\delta_{(3,-4)}
=0.
\end{equation*} 
In the case of no further interactions the system of the four pulses is 
$2$-closed if the nonresonance conditions
\begin{alignat*}{4}
\delta_{(1,1)}
&\neq0,\qquad
&\delta_{(2,2)}=\delta_{(-1,4)}
&\neq0,\qquad
&\delta_{(3,3)}=\delta_{(1,4)}
&\neq0,\qquad
&\delta_{(4,4)}
&\neq0,
\\
\delta_{(1,-2)}
&\neq0,\qquad
&\delta_{(1,3)}
&\neq0,\qquad
&\delta_{(2,4)}
&\neq0,\qquad
&\delta_{(3,4)}
&\neq0
\end{alignat*}
%
are satisfied. 
The equations \eqref{macro1} yield
\begin{equation*}
\begin{cases}
\ds \partial_\tau A_{1,1}-\nabla_\vth\Omega(\vth_1){\cdot}\nabla_y A_{1,1}
=-\i\frac{\ol{c_{(1,2)}}}{\Omega(\vth_1)}\ol{A_{1,2}}A_{1,3},
\\[5mm]
\ds \partial_\tau A_{1,2}-\nabla_\vth\Omega(\vth_2){\cdot}\nabla_y A_{1,2}
=-\i\frac1{\Omega(\vth_2)}
\big(\ol{c_{(1,2)}}\ol{A_{1,1}}A_{1,3}+\ol{c_{(2,3)}}\ol{A_{1,3}}A_{1,4}\big),
\\[5mm]
\ds \partial_\tau A_{1,3}-\nabla_\vth\Omega(\vth_3){\cdot}\nabla_y A_{1,3}
=-\i\frac1{\Omega(\vth_3)}
\big(c_{(1,2)}A_{1,1}A_{1,2}+\ol{c_{(2,3)}}\ol{A_{1,2}}A_{1,4}\big),
\\[5mm]
\ds \partial_\tau A_{1,4}-\nabla_\vth\Omega(\vth_4){\cdot}\nabla_y A_{1,4}
=-\i\frac{c_{(2,3)}}{\Omega(\vth_4)}A_{1,2}A_{1,3}
\end{cases}
\end{equation*}
with $\vth_3=\vth_1+\vth_2$, $\Om(\vth_3)=\Om(\vth_1)+\Om(\vth_2)$,
$\vth_4=\vth_2+\vth_3$, $\Om(\vth_4)=\Om(\vth_2)+\Om(\vth_3)$,
and
$c_{(1,2)}$ given by \eqref{c12},
\begin{align*}
 c_{(2,3)}&=
 -4\i\sum_{\alpha\in\Gamma}a_{2,\alpha}
 \sin\left(\frac{\vth_2}2{\cdot}\alpha\right)
 \sin\left(\frac{\vth_3}2{\cdot}\alpha\right)
 \sin\left(\frac{
 \vth_4
 }2{\cdot}\alpha\right)
 -b_2.
\end{align*}
As in the previous sections, the coefficients $A_{2,(j,i)}$ 
with $(j,i)$ as in the nonresonance conditions 
can be calculated by the equations \eqref{macro2},
while the coefficients $A_{2,(j,i)}$ 
with $(j,i)$ as in the resonance conditions are included in 
the second-order amplitudes $A_{2,j}$, $j=1,\ldots,4$,
and thus remain undetermined.

Note, how the right hand sides of the amplitude equations mirror the two
three-wave-interactions.
In particular, setting identically $A_{1,4}=0$, 
the three first equations give the three-wave-interaction equations 
\eqref{3wi}, 
and the fourth equation becomes $A_{1,2}A_{1,3}=0$.
%
As discussed previously 
(cf.\ Sec.\ \ref{threepulsesclosedN2}, Case \ref{case3wi} and 
Sec.\ \ref{twopulsesnonclosedN2}, Case 1),
the latter condition obstructs the reasonable study of the dynamics of the
system (already by restricting the choice of initial data)
and is safely guaranteed if one of the two amplitudes vanishes.
However, this leads subsequently to the cases cited above, reducing
eventually 
the system for the dynamics of four pulses to a single transport
equation for one of them.
%
%

%
%


\subsubsection{3-closed system of a three-wave-interaction} 
\label{433}
We conclude our list of examples by calculating 
the third-order approximation $X_{3}^{A,\ve}$ 
for a set of three different pulses $\{\pm(\vth_j,\om_j):\ j=1,2,3\}$, 
$\om_j=\Om(\vth_j)>0$, 
which satisfy the resonance condition \eqref{ca2rescond} 
of Case \ref{case3wi} in Sec.\ \ref{threepulsesclosedN2}.
We assume that except for this three-wave-interaction,
$(\vth_1{+}\vth_2,\om_1{+}\om_2)=(\vth_3,\om_3)$,
no other interaction (of order $2$) takes place among the three pulses.
Moreover, in order to be able to calculate $X_{3}^{A,\ve}$ such that 
$\res\big(X_{3}^{A,\ve}\big)=\OO(\ve^4)$,
we assume that the set is closed under interactions up to order $3$.
According to Definition \ref{defclosure}, in the present setting
this is the case when the nonresonance conditions 
(or order $2$) 
\eqref{nu3N2ca2clcond} 
are satisfied and additionally
\begin{center}
\emph{
the pairs
$(\kappa\vth_1+\lambda\vth_2,\kappa\om_1+\lambda\om_2)$ 
with $(\kappa,\lambda)=
(3,0),\ (0,3),\ (3,3),\ (2,{-}1),\ ({-}1,2),\ (3,1),\ (1,3),\ (3,2),\ (2,3)
$
\\do not characterize pulses, except if 
either $(3\vth_1,3\om_1)=(\vth_2,\om_2)$ or $(3\vth_2,3\om_2)=(\vth_1,\om_1)$.
}\end{center}
%
%
Hence, excluding the latter exceptional cases (resonances of order $3$ within
the set of pulses), 
the system is closed under interactions up to order $3$ 
if in addition to \eqref{nu3N2ca2clcond} the 
nonresonance conditions (of order $3$)
%
\begin{align} 
&\delta_{(j,j,j)}\neq0,\quad j=1,2,3,\qquad
\delta_{(1,1,-2)}\neq0,\qquad
\delta_{(2,2,-1)}\neq0,
\notag\\
& \delta_{(1,1,3)}\neq0,\qquad 
\delta_{(2,2,3)}\neq0,\qquad 
\delta_{(3,3,1)}\neq0,\qquad
\delta_{(3,3,2)}\neq0.
\label{nu3N3ca2clcond}
\end{align}
are satisfied.
Assuming that all these nonresonance conditions 
as well as those in \eqref{nu3N2ca2clcond} are mutually independent,
the third-order approximation reads
\begin{align}
&X_3^{A,\ve}=X_2^{A,\ve}+\ve^{3}\Bigg(
\frac12A_{3,(1,-1)}+
\sum_{j=1}^3 \Big(A_{3,j}\boe_{j}
+A_{3,(j,j)}
\boe_j^2
+A_{3,(j,j,j)}
\boe_j^3
\Big)
\notag\\&
+A_{3,(1,3)}
\boe_1 \boe_3 
+A_{3,(1,-2)}
\boe_1 \boe_{-2} 
+A_{3,(2,3)}
\boe_2\boe_3 
+A_{3,(1,1,-2)}
\boe_1^2 \boe_{-2} 
+A_{3,(2,2,-1)}
\boe_2^2\boe_{-1} 
\notag\\&
+A_{3,(1,1,3)}
\boe_1^2 \boe_3 
+A_{3,(2,2,3)}
\boe_2^2 \boe_3 
+A_{3,(3,3,1)}
\boe_3^2\boe_1
+A_{3,(3,3,2)}
\boe_3^2 \boe_2 
+\cc
\Bigg)
\label{nu3N3ca2}
\end{align}
with $X_2^{A,\ve}$ given by \eqref{nu3N2inter} in  
Sec.\ \ref{threepulsesclosedN2}, Case \ref{case3wi}.
There, we determined for $X_2^{A,\ve}$
the first-order amplitudes $A_{1,j}$ 
(as solutions of the three-wave-interaction equations \eqref{3wi})
and the functions \eqref{3wimixed}.
%
Here, using \eqref{3wimixed},
we obtain from the equations \eqref{macro3} for $k=3$
the following system of linearly coupled inhomogeneous transport equations 
for the second-order amplitudes $A_{2,j}$:
 \begin{align*}
 &\partial_\tau A_{2,1}
 -\nabla_\vth\Omega(\vth_1){\cdot}\nabla_y A_{2,1}
 +\i\frac{\ol{c_{(1,2)}}}{\Omega(\vth_1)}
 \big(A_{1,3}\ol{A_{2,2}}+\ol{A_{1,2}}A_{2,3}\big)
 \\&\quad
 =
 \frac1{2\i\Omega(\vth_1)}
 \Big(\big(
 (\eta_{(1,1)}{+}{\textstyle\frac{2b_2^2}{b_1}})|A_{1,1}|^2
 +2\eta_{(1,-2)}|A_{1,2}|^2 
 +2\eta_{(1,3)}|A_{1,3}|^2 
 \big)A_{1,1} 
 -\partial_\tau^2 A_{1,1}
 \\&\qquad\qquad\qquad\
 + A_{1,3}\,\gamma_{(3,-2)}{\cdot}\nabla_y \ol{A_{1,2}}
 +\ol{A_{1,2}}\,\gamma_{(-2,3)}{\cdot}\nabla_y A_{1,3}
 +\frac12\sum_{\alpha\in\Gamma}a_{1,\alpha}
 \e^{\i\vth_1\cdot\alpha}(\alpha{\cdot}\nabla_y)^2 A_{1,1}
 \Big),
 \displaybreak[0]\\&
 \partial_\tau A_{2,2}
 -\nabla_\vth\Omega(\vth_2){\cdot}\nabla_y A_{2,2}
 +\i\frac{\ol{c_{(1,2)}}}{\Omega(\vth_2)}
 \big(A_{1,3}\ol{A_{2,1}}+\ol{A_{1,1}}A_{2,3}\big)
 \\&\quad
 =
 \frac1{2\i\Omega(\vth_2)}
 \Big(\big(
 (\eta_{(2,2)}{+}{\textstyle\frac{2b_2^2}{b_1}})
 |A_{1,2}|^2
 +2\eta_{(2,-1)}|A_{1,1}|^2
 +2\eta_{(2,3)}|A_{1,3}|^2
 \big)A_{1,2}
 -\partial_\tau^2 A_{1,2}
 \\&\qquad\qquad\qquad\ 
 +A_{1,3}\,\gamma_{(3,-1)}{\cdot}\nabla_y\ol{A_{1,1}}
 +\ol{A_{1,1}}\,\gamma_{(-1,3)}{\cdot}\nabla_y A_{1,3}
 +\frac12\sum_{\alpha\in\Gamma}a_{1,\alpha}
 \e^{\i\vth_2\cdot\alpha}(\alpha{\cdot}\nabla_y)^2 A_{1,2}
 \Big),
 \displaybreak[0]\\&
 \partial_\tau A_{2,3}
 -\nabla_\vth\Omega(\vth_3){\cdot}\nabla_y A_{2,3}
 +\i\frac{c_{(1,2)}}{\Omega(\vth_3)}
 \big(A_{1,1}A_{2,2}+A_{1,2}A_{2,1}\big)
 \\&\quad
 =
 \frac1{2\i\Omega(\vth_3)}
 \Big(\big(
 (\eta_{(3,3)}{+}{\textstyle\frac{2b_2^2}{b_1}})
 |A_{1,3}|^2
 +2\eta_{(3,1)}|A_{1,1}|^2
 +2\eta_{(3,2)}|A_{1,2}|^2
 \big)A_{1,3}
 -\partial_\tau^2 A_{1,3}
 \\&\qquad\qquad\qquad\
 +A_{1,1}\,\gamma_{(1,2)}{\cdot}\nabla_y A_{1,2}
 +A_{1,2}\,\gamma_{(2,1)}{\cdot}\nabla_y A_{1,1}
 +\frac12\sum_{\alpha\in\Gamma}a_{1,\alpha}
 \e^{\i\vth_3\cdot\alpha}(\alpha{\cdot}\nabla_y)^2 A_{1,3}
 \Big)
 \end{align*}
with 
$c_{(1,2)}$ given by \eqref{c12}, and 
\begin{align*}
\gamma_{(p,q)}
&:=2\sum_{\alpha\in\Gamma}a_{2,\alpha}
  (\cos((\vth_p{+}\vth_q){\cdot}\alpha){-}\cos(\vth_q{\cdot}\alpha))\alpha,
\quad 
\eta_{(j,p)}
:=
\frac{2b_2^2}{b_1}
 +2 \frac{|c_{(j,p)}|^2}{\delta_{(j,p)}}
 +3 c_{(j,p,-p)}.
\end{align*}
%
(Note, that the source terms and the coefficients 
in front 
of the amplitudes $A_{2,j}$ are known.)
%
%
%
Determining $A_{2,j}$ by these equations, 
and since the set of pulses is $3$-closed, 
we can then calculate
by \eqref{macro4} for $k=3$ the remaining functions of $X_3^{A,\ve}$,
except for the third-order amplitudes $A_{3,j}$.
As previously, setting e.g. $A_{3,j}=0$, we obtain 
$\res\big(X_3^{A,\ve}\big)=\OO(\ve^4)$.
%
\subsection{Existence of interacting pulses}
\label{interactionexistence}
We conclude this section of examples by discussing the question 
whether there really exist interacting pulses in lattices.
We do this by showing exemplarily 
the existence of a closed system of three different pulses
$\{\pm(\vth_j,\om_j): j=1,2,3\}$
which satisfy the resonance condition \eqref{ca2rescond} 
(three-wave-interaction)
and the nonresonance conditions \eqref{nu3N2ca2clcond}
of Case~\ref{case3wi} in Sec.\ \ref{threepulsesclosedN2}
for a one-dimensional lattice (i.e., a chain) of the form 
\eqref{M} with $d=1$, $\Gamma=\Z$ and 
only a nearest-neighbour interaction potential 
$V_1(x)=\frac{a_1}2 x^2+O(|x|^3)$, 
where $a_1:=a_{1,1}$ in \eqref{VW} 
and $V_\alpha=0$ for $\alpha\in\N\setminus\{1\}$.
The linearized model and the dispersion function read, respectively,
\begin{equation*}
\ddot x_\g=a_1(x_{\g+1}-2x_\g+x_{\g-1})-b_1 x_\g,\ \g\in\Z,
\quad\text{and}\quad
\Om(\vth)=\sqrt{2a_1(1-\cos\vth)+b_1}>0,
\end{equation*}
where 
the stability condition \eqref{SC} is satisfied iff 
we assume $\min\{b_1,b_1+4a_1\}>0$.

For three given pulses $\pm(\vth_j,\om_j)$, $j=1,2,3$, 
with $\om_j:=\Om(\vth_j)>0$ 
the resonance condition \eqref{ca2rescond} is satisfied
if and only if the wave numbers $\vth_1,\vth_2\in {\mathbb T}_\Z$ 
solve the equation 
\begin{equation}\label{omegaka}
\Om^2(k_1\vth_1{+}k_2\vth_2)=\big(k_1\Om(\vth_1)+k_2\Om(\vth_2)\big)^2,
\quad (k_1,k_2)\in\Z^2
\end{equation}
for $(k_1,k_2)=(1,1)$.
Moreover, the nonresonance conditions \eqref{nu3N2ca2clcond} are satisfied 
if and only if (the same) 
$\vth_1,\vth_2$ do \emph{not} 
solve the equations \eqref{omegaka} for
\begin{equation}\label{kas}
(k_1,k_2)=(2,0),\ (0,2),\ (2,2),\ (2,1),\ (1,2),\ (1,-1).
\end{equation}
Inserting the dispersion function $\Om(\vth)$, equation \eqref{omegaka} 
becomes 
\begin{align*}
-\sign(a_1)
\cos(k_1\vth_1{+}k_2\vth_2)
&=(k_1^2{+}k_2^2{-}1)f
-\sign(a_1)(k_1^2\cos\vth_1+k_2^2\cos\vth_2)
\\&\quad\ 
+2k_1k_2\sqrt{(f{-}\sign(a_1)\cos\vth_1)(f{-}\sign(a_1)\cos\vth_2)}
\end{align*}
with $\ds f:
=\frac{b_1}{2|a_1|}+\sign(a_1)
$.
In particular, for $(k_1,k_2)=(1,1)$, 
applying the trigonometric theorem and making the substitutions
$\ds \chi:=\frac{1-\cos\vth_1}2,\ \psi:=\frac{1-\cos\vth_2}2\in[0,1]$, 
the equation reads 
\begin{equation*}
\sign(a_1)s\sqrt{\chi(1{-}\chi)\psi(1{-}\psi)}
=\frac\phi2+\sign(a_1)\chi\psi
+\sqrt{(\phi{+}\sign(a_1)\chi)(\phi{+}\sign(a_1)\psi)}
\end{equation*}
with $s:=\sign(\sin\vth_1\sin\vth_2)$ and $\ds\phi:
=\frac{b_1}{4|a_1|}
>0$.
Squaring the left and right hand sides of this equation,
we obtain
\begin{equation*}
-\chi\psi(\chi+\psi)
=\frac{5\phi^2}4+\sign(a_1)\phi(\chi\psi+\chi+\psi)
+(\phi+2\sign(a_1)\chi\psi)\sqrt{(\phi{+}\sign(a_1)\chi)(\phi{+}\sign(a_1)\psi)}.
\end{equation*}
Obviously, for $a_1>0$ there exist no solutions $\chi,\psi\in[0,1]$ to this
equation.
However, for $a_1\in(-b_1/4,0)$ the stability condition 
\eqref{SC} 
still holds, 
and the 
equation reads
\begin{align*}
g(\chi,\psi):=\frac{5\phi^2}4+\chi\psi(\chi+\psi)-\phi(\chi\psi+\chi+\psi)
+(\phi-2\chi\psi)\sqrt{(\phi-\chi)(\phi-\psi)}=0
\end{align*}
with $\phi>1$.
The function $g$ is symmetric 
and for
\begin{equation*}
\tilde g(\chi):=g(\chi,\chi)=
4\chi^3-3\phi\chi^2-3\phi\chi+9\phi^2/4
\end{equation*}
we get $\tilde g(0)=9\phi^2/4$, $\tilde g(1)=9(\phi-4/3)^2/4$, 
$\tilde g^\prime(\chi)=3(4\chi^2-2\phi\chi-\phi)$,
$\tilde g^{\prime\prime}(\chi)=6(4\chi-\phi)$.
Thus, $\tilde g$ has a minimum at $\chi_m(\phi):=(\phi/4)(1+\sqrt{1+4/\phi})$,
which lies in $(0,1)$ if and only if $\phi<4/3$, 
and 
$\tilde g(\chi_m(\phi))=-(\phi^3/8)[1-12/\phi+(1+4/\phi)^{3/2})]<0$, 
since  
for $x:=4/\phi>3$ 
we have
$(1+x)^{3/2}>3x-1 \Leftrightarrow (1+x)^3>(3x-1)^2 \Leftrightarrow x(x-3)^2>0$.

Hence, for $\phi\in(1,4/3)$, i.e. for $a_1\in(-b_1/4,-3b_1/16)$
the level set 
$\{(\chi,\psi)\in[0,1]^2: g(\chi,\psi)=0\}$
is nonempty, and its elements solve 
\begin{equation}\label{rcchain}
-s\sqrt{\chi(1{-}\chi)\psi(1{-}\psi)}
=\frac\phi2-\chi\psi+\sqrt{(\phi-\chi)(\phi-\psi)}
\end{equation}
with the $s=\pm1$ chosen according to the sign of the right hand side.
Changing if necessary the sign of $\vth_1$ or $\vth_2$ 
in such a way that $s=\sign(\sin\vth_1\sin\vth_2)$,
we thus obtain a whole family of $(\vth_1,\vth_2)\in {\mathbb T}_\Z^2$,
which satisfy the resonance condition \eqref{rcchain}
or, equivalently,
\begin{equation*}
\cos(\vth_1{+}\vth_2)
=f+\cos\vth_1+\cos\vth_2
+2\sqrt{(f+\cos\vth_1)(f+\cos\vth_2)}
\end{equation*}
with $\ds f=2\phi-1>1$. 
Among the family of solutions $(\vth_1,\vth_2)\in {\mathbb T}_\Z^2$ satisfying 
\eqref{rcchain} we then have to find a pair which
satisfies also the nonresonance conditions 
\begin{equation*}
\cos(k_1\vth_1{+}k_2\vth_2)
\neq
(k_1^2{+}k_2^2{-}1)f+k_1^2\cos\vth_1+k_2^2\cos\vth_2
+2k_1k_2\sqrt{(f{+}\cos\vth_1)(f{+}\cos\vth_2)}
\end{equation*}
for the values of $(k_1,k_2)$ given in \eqref{kas}.
This is equivalent to finding 
$(\chi,\psi)\in\{(\chi,\psi)\in[0,1]^2: g(\chi,\psi)=0\}$
which satisfy 
\begin{align*}
\chi^2,\ \psi^2,\ \zeta^2\neq \frac{3\phi}4,\quad
\chi\psi,\ \chi\zeta,\ \zeta\psi\neq\frac\phi2
\quad\text{with}\quad\zeta:=
\chi+\psi-\phi-2\sqrt{(\phi-\chi)(\phi-\psi)}.
 \end{align*}
The intersection of the corresponding curves in $[0,1]^2$ 
with the curve of solutions is a zero-measure set.
Hence, there exist uncountably many   
$(\vth_1,\vth_2)\in {\mathbb T}_\Z^2$ which 
satify \eqref{omegaka} with $(k_1,k_2)=(1,1)$ (resonance condition)
and do not satisfy \eqref{omegaka} with the $(k_1,k_2)$ given in \eqref{kas}
(i.e., satisfy the nonresonance conditions).\\ 

%
%
%
%
\noindent {\bf Acknowledgments:} 
This work has been partially supported by the DFG Priority Program 1095 
{\it Analysis, Modeling and Simulation of Multiscale Problems}  
under 
Mi 459/3--3. 
I thank {\sc Alexander Mielke}, 
who suggested this idea to me,
and {\sc Michael Herrmann}
for the many fruitful discussions. 
%
%
%
%
\renewcommand{\baselinestretch}{0.92} \small 
%
%

\end{document}